\documentclass[sigconf,authorversion]{acmart}
\usepackage{tabu}                      % only used for the table example
\usepackage{booktabs}                  % only used for the table example
\usepackage{lipsum}                    % used to generate placeholder text
\usepackage{mwe}                       % used to generate placeholder figures
\usepackage{caption}
\usepackage{colortbl}
\usepackage{makecell}
\usepackage{float}
\hypersetup{
    colorlinks=true,
    linkcolor=blue,
    filecolor=magenta,      
    urlcolor=cyan,
    pdftitle={Overleaf Example},
    }

\usepackage{xcolor}

\usepackage{tikz}
\def\\checkmark{\tikz\fill[scale=0.4](0,.35) -- (.25,0) -- (1,.7) -- (.25,.15) -- cycle;} 
\newcolumntype{P}[1]{>{\centering\arraybackslash}p{#1}}
\AtBeginDocument{%
  \providecommand\BibTeX{{%
    \normalfont B\kern-0.5em{\scshape i\kern-0.25em b}\kern-0.8em\TeX}}}

\acmDOI{10.1145/3706598.3713566}
\begin{document}

\title{The Fidelity-based Presence Scale (FPS): Modeling the Effects of Fidelity on Sense of Presence \\ (To appear in CHI 2025)}

\author{Jacob Belga}
\affiliation{%
  \institution{University of Central Florida}
  \city{Orlando}
  \state{Florida}
  \country{USA}
}
\email{jacob.belga@ucf.edu}

\author{Richard Skarbez}
\affiliation{%
  \institution{La Trobe University}
  \city{Melbourne}
  \state{Victoria}
  \country{AUS}
}
\email{r.skarbez@latrobe.edu.au}

\author{Yahya Hmaiti}
\affiliation{%
  \institution{University of Central Florida}
  \city{Orlando}
  \state{Florida}
  \country{USA}
}
\email{Yohan.Hmaiti@ucf.edu}

\author{Eric J. Chen}
\affiliation{%
  \institution{Seminole High School}
  \city{Sanford}
  \state{Florida}
  \country{USA}
}
\email{ezhu2006@yahoo.com}

\author{Ryan P. McMahan}
\affiliation{%
  \institution{Virginia Tech}
  \city{Blacksburg}
  \state{Virginia}
  \country{USA}
}
\email{rpm@vt.edu}

\author{Joseph J. LaViola Jr.}
\affiliation{%
  \institution{University of Central Florida}
  \city{Orlando}
  \state{Florida}
  \country{USA}
}
\email{jlaviola@ucf.edu}

\newcommand{\ReviewerOne}[1]{\textcolor{teal}{\textbf{} #1}}
\newcommand{\ReviewerTwo}[1]{\textcolor{teal}{\textbf{} #1}}
\newcommand{\ReviewerThree}[1]{\textcolor{teal}{\textbf{} #1}}
\newcommand\redsout{\bgroup\markoverwith{\textcolor{red}{\rule[0.5ex]{2pt}{0.4pt}}}\ULon}

\renewcommand{\shortauthors}{Belga et al.}

\begin{abstract}
Within the virtual reality (VR) research community, there have been several efforts to develop questionnaires with the aim of better understanding the sense of presence. Despite having numerous surveys, the community does not have a questionnaire that informs which components of a VR application contributed to the sense of presence. Furthermore, previous literature notes the absence of consensus on which questionnaire or questions should be used. Therefore, we conducted a Delphi study, engaging presence experts to establish a consensus on the most important presence questions and their respective verbiage. We then conducted a validation study with an exploratory factor analysis (EFA). The efforts between our two studies led to the creation of the Fidelity-based Presence Scale (FPS). With our consensus-driven approach and fidelity-based factoring, we hope the FPS will enable better communication within the research community and yield important future results regarding the relationship between VR system fidelity and presence.
\end{abstract}





\begin{teaserfigure}
    \includegraphics[width=\textwidth]{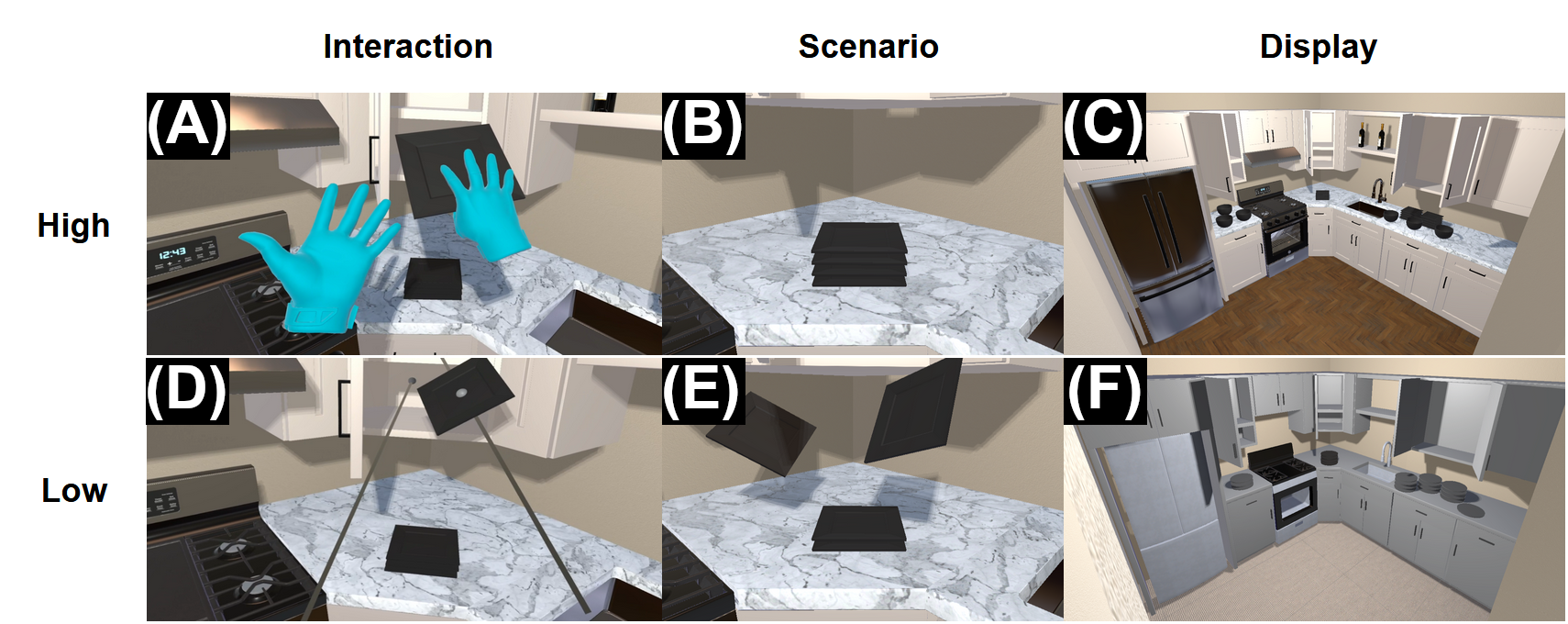}
    \captionsetup{justification=centering}
    \caption{The system fidelity aspects we implemented in our validation study. The top row (A-C) shows the high-fidelity versions of the interaction, scenario, and display components of our virtual reality (VR) study. The bottom row (D-F) shows the low-fidelity versions: D) The Fishing Reel technique \cite{bowman1997evaluation} is used for interactions instead of the Virtual Hand technique \cite{poupyrev1998egocentric}; E) Gravity is not consistent within the scenario; and F) Low-poly models and basic lighting are used instead of high-poly models and advanced lighting.}
    \Description{The system fidelity aspects we implemented in our validation study. The top row (A-C) shows the high-fidelity versions of the interaction, scenario, and display components of our virtual reality (VR) study. The bottom row (D-F) shows the low-fidelity versions: D) The Fishing Reel technique \cite{bowman1997evaluation} is used for interactions instead of the Virtual Hand technique \cite{poupyrev1998egocentric}; E) Gravity is not consistent within the scenario; and F) Low-poly models and basic lighting are used instead of high-poly models and advanced lighting.}
    \label{fig:TeaserFigure}
\end{teaserfigure}

\maketitle
\section{Introduction}

\textit{Presence}, sometimes noted as ``spatial presence'', is a widely investigated sensation in virtual reality (VR) research and is defined as the sense of ``being there'' in a VR environment \cite{skarbez2017survey}. Since its conception in the 1980s \cite{minsky1980telepresence}, numerous techniques have been produced in an effort to better understand and measure presence. Physiological measures, behavioral measures, semi-structured interviews, and questionnaires are some of the common employed techniques found throughout the literature \cite{tran2024survey}. Of these techniques, questionnaires are the most widely used in user studies. One potential reason for the popularity of questionnaires is that they afford researchers the ability to replicate studies \cite{echtler2018open}. Previous literature reviews \cite{souza2021measuring, skarbez2017survey, van2004measuring} have shown that a multitude of presence questionnaires have been developed and used. Despite the numerous questionnaire options, many researchers still find it necessary to modify and customize these questionnaires \cite{tran2024survey}. This is concerning as altering the structure of a validated presence survey can lead to adverse results and diminishes the ability to generalize results among multiple studies \cite{laviola20173d}. This indicates that there is no single presence questionnaire that addresses the broader needs of the research community.

One possible reason for the absence of a presence instrument that meets the needs of the community is that most presence questionnaires do not account for the realism or fidelity of a VR application \cite{skarbez2017survey}. Instead, current presence questionnaires provide insight to the user's general response of being in a virtual environment. For example, the Igroup Presence Questionnaire (IPQ) employs the sub-scales of Presence, Spatial Presence, Involvement, and Realness \cite{schubert2001experience}. As another example, the Spatial Presence Experience Scale (SPES) provides self-location and possible actions as their sub-scales \cite{hartmann2015spatial}. We believe that VR researchers and developers would greatly benefit from knowing which aspects of a VR application and their respective fidelity impact a user's sense of presence.

In this paper, we present the development and validation of a presence questionnaire, co-designed by the broader community, to assess how different aspects of fidelity impact presence. We used McMahan et al.'s \cite{McMahan2018, mcmahan2016affect} \textit{system fidelity} framework, which encompasses interaction fidelity (e.g. object manipulation, locomotion technique), scenario fidelity (e.g. gravity of objects, behavior of agents/objects) and display fidelity (e.g. polygon count, audio quality), as our theoretical foundation to address the different aspects of fidelity and selected preexisting presence questionnaires encompassing these aspects. System fidelity is defined as the objective degree in which real-world interactions or experiences are replicated by an interactive system \cite{McMahan2018, mcmahan2016affect}. We then conducted a Delphi study \cite{hsu2007delphi, keeney2006consulting} with 16 presence experts from the broader research community to identify the most important presence questions and what aspects of fidelity they are most affected by. Our Delphi study yielded an initial presence survey comprising of 11 items.

We also present a $2\times2\times2$ within-subject study that controlled interaction, scenario, and display fidelity at low and high levels (see Figure \ref{fig:TeaserFigure}). We used the results of this study (n=55) to conduct an exploratory factor analysis (EFA) to identify the new structure of our presence survey, which yielded 10 items across three factors: Interaction Presence, Scenario Presence, and Display Presence. We then present the results of our experiment, which confirm that our new presence survey is capable of identifying how each fidelity component affects presence. Between our Delphi study and validation study, our efforts have culminated in the development of the Fidelity-based Presence Scale (FPS).
In summary, our contributions are:
\begin{itemize}
    \item Co-designing a presence questionnaire with perspectives of expert presence researchers from the broader community.
    \item Conducting one of the first studies to investigate the effects of interaction, scenario, and display fidelity on one's sense of presence.
    \item Using EFA to validate the reliability of our new presence questionnaire and to identify its three subscales: Interaction Presence, Scenario Presence, and Display Presence
    \item Providing results that demonstrate the validity of the FPS and its ability to indicate how different aspects of VR system fidelity affect presence.
    \item Providing open access to FPS in common survey formats (e.g., PDF, Word, Qualtrics): \href{https://osf.io/r78m4/?view_only=ff102c8ac9914fae8d7a6495bc3d627d}{\textcolor{blue}{FPS Survey}}.
\end{itemize}

\section{Related Work}

From our review of presence literature, which builds upon systematic surveys from Van Baren \cite{van2004measuring}, Skarbez et al. \cite{skarbez2017survey}, and Souza et al. \cite{souza2021measuring}, we have learned of the broader perspectives of presence and the current measurement tools used to evaluate presence. In our Delphi study, we include a series of questionnaires that are predominantly utilized in VR contexts across the presence research community and capture nuances from the past 40 years of research. Table \ref{tab:RelatedWork} presents a series of presence questionnaires along with their respective item counts, measurement factors, whether they conducted an EFA, and whether they conducted a Delphi study.

\begin{table*}[t]
\scriptsize
\centering
\begin{tabular}{ccccccc}
\rowcolor[HTML]{FFFFFF} 
{\color[HTML]{000000} \textbf{Citation}} &
  {\color[HTML]{000000} \textbf{Questionnaire}} &
  {\color[HTML]{000000} \textbf{Items}} &
  {\color[HTML]{000000} \textbf{Year}} &
  {\color[HTML]{000000} \textbf{\begin{tabular}[c]{@{}c@{}}Measurement\\ Factors\end{tabular}}} &
  {\color[HTML]{000000} \textbf{\begin{tabular}[c]{@{}c@{}}Delphi \\ Study\end{tabular}}} &
  {\color[HTML]{000000} \textbf{\begin{tabular}[c]{@{}c@{}}Exploratory \\ Factor\\ Analysis\end{tabular}}} \\ \hline
\rowcolor[HTML]{EFEFEF} 
\cite{barfield1993sense} &
  Barfield et al. Questionnaire &
  3 &
  1993 &
  \multicolumn{1}{c|}{\cellcolor[HTML]{EFEFEF}N/A} &
   &
   \\
\rowcolor[HTML]{FFFFFF} 
{\color[HTML]{000000} \cite{kim1997telepresence}} &
  {\color[HTML]{000000} Kim \& Biocca Questionnaire} &
  {\color[HTML]{000000} 8} &
  {\color[HTML]{000000} 1997} &
  \multicolumn{1}{c|}{\cellcolor[HTML]{FFFFFF}{\color[HTML]{000000} Departure, Arrival}} &
  {\color[HTML]{000000} } &
  {\color[HTML]{000000} \checkmark} \\
\rowcolor[HTML]{EFEFEF} 
\cite{parent1998virtual} &
  \begin{tabular}[c]{@{}c@{}}Questionnaire on Presence\\ and Realism\end{tabular} &
  10 &
  1998 &
  \multicolumn{1}{c|}{\cellcolor[HTML]{EFEFEF}N/A} &
   &
   \\
\rowcolor[HTML]{FFFFFF} 
{\color[HTML]{000000} \cite{witmer1998measuring}} &
  {\color[HTML]{000000} \textbf{\begin{tabular}[c]{@{}c@{}}Witmer-Singer\\ Presence Questionnaire\end{tabular}}} &
  {\color[HTML]{000000} 32} &
  {\color[HTML]{000000} 1998} &
  \multicolumn{1}{c|}{\cellcolor[HTML]{FFFFFF}{\color[HTML]{000000} \begin{tabular}[c]{@{}c@{}}Involved/Control, Natural, Auditory,\\ Haptic, Resolution, Interface Quality\end{tabular}}} &
  {\color[HTML]{000000} } &
  {\color[HTML]{000000} } \\
\rowcolor[HTML]{EFEFEF} 
\cite{dinh1999evaluating} &
  Dinh et al. Questionnaire &
  14 &
  1999 &
  \multicolumn{1}{c|}{\cellcolor[HTML]{EFEFEF}N/A} &
   &
   \\
\rowcolor[HTML]{FFFFFF} 
{\color[HTML]{000000} \cite{lombard2000measuring}} &
  {\color[HTML]{000000} \begin{tabular}[c]{@{}c@{}}Lombard \& Ditton\\ Questionnaire\end{tabular}} &
  {\color[HTML]{000000} 103} &
  {\color[HTML]{000000} 2000} &
  \multicolumn{1}{c|}{\cellcolor[HTML]{FFFFFF}{\color[HTML]{000000} \begin{tabular}[c]{@{}c@{}}Immersion, Parasocial interaction, \\ Parasocial relationships,\\ Physiological responses, Social reality, \\ Interpersonal social richness, \\ General social richness\end{tabular}}} &
  {\color[HTML]{000000} } &
  {\color[HTML]{000000} \checkmark} \\
\rowcolor[HTML]{EFEFEF} 
\cite{murray2000presence} &
  Murray et al. Questionnaire &
  5 &
  2000 &
  \multicolumn{1}{c|}{\cellcolor[HTML]{EFEFEF}N/A} &
   &
   \\
\rowcolor[HTML]{FFFFFF} 
{\color[HTML]{000000} \cite{nichols2000measurement}} &
  {\color[HTML]{000000} Nichols et al. Questionnaire} &
  {\color[HTML]{000000} 9} &
  {\color[HTML]{000000} 2000} &
  \multicolumn{1}{c|}{\cellcolor[HTML]{FFFFFF}{\color[HTML]{000000} N/A}} &
  {\color[HTML]{000000} } &
  {\color[HTML]{000000} } \\
\rowcolor[HTML]{EFEFEF} 
\cite{banos2000presence} &
  \begin{tabular}[c]{@{}c@{}}Reality Judgment and Presence\\ Questionnaire\end{tabular} &
  18 &
  2000 &
  \multicolumn{1}{c|}{\cellcolor[HTML]{EFEFEF}\begin{tabular}[c]{@{}c@{}}Reality Judgement, Internal/External Correspondence,\\ Attention/Absorption\end{tabular}} &
   &
  \checkmark \\
\rowcolor[HTML]{FFFFFF} 
{\color[HTML]{000000} \cite{usoh2000using}} &
  {\color[HTML]{000000} \textbf{\begin{tabular}[c]{@{}c@{}}Slater-Usoh-Steed Presence\\ Questionnaire\end{tabular}}} &
  {\color[HTML]{000000} 6} &
  {\color[HTML]{000000} 2000} &
  \multicolumn{1}{c|}{\cellcolor[HTML]{FFFFFF}{\color[HTML]{000000} N/A}} &
  {\color[HTML]{000000} } &
  {\color[HTML]{000000} \textbf{}} \\
\rowcolor[HTML]{EFEFEF} 
\cite{gerhard2001continuous} &
  Gerhard et al. Questionnaire &
  19 &
  2001 &
  \multicolumn{1}{c|}{\cellcolor[HTML]{EFEFEF}Immersion, Communication, Involvement, Awareness} &
   &
   \\
\rowcolor[HTML]{FFFFFF} 
{\color[HTML]{000000} \cite{schubert2001experience}} &
  {\color[HTML]{000000} \textbf{\begin{tabular}[c]{@{}c@{}}Igroup Presence\\ Questionnaire\end{tabular}}} &
  {\color[HTML]{000000} 14} &
  {\color[HTML]{000000} 2001} &
  \multicolumn{1}{c|}{\cellcolor[HTML]{FFFFFF}{\color[HTML]{000000} \begin{tabular}[c]{@{}c@{}}Presence (PRES), Spatial Presence (SP),\\ Involvement (INV), Realness (REAL\end{tabular}}} &
  {\color[HTML]{000000} } &
  {\color[HTML]{000000} \checkmark} \\
\rowcolor[HTML]{EFEFEF} 
\cite{lessiter2001cross} &
  \begin{tabular}[c]{@{}c@{}}ITC Sense of Presence\\ Inventory\end{tabular} &
  63 &
  2001 &
  \multicolumn{1}{c|}{\cellcolor[HTML]{EFEFEF}\begin{tabular}[c]{@{}c@{}}Sense of Physical Space, Engagement, Ecological Validity,\\ Negative Effects\end{tabular}} &
   &
  \checkmark \\
\rowcolor[HTML]{FFFFFF} 
{\color[HTML]{000000} \cite{krauss2001measurement}} &
  {\color[HTML]{000000} Krauss et al. Questionnaire} &
  {\color[HTML]{000000} 42} &
  {\color[HTML]{000000} 2001} &
  \multicolumn{1}{c|}{\cellcolor[HTML]{FFFFFF}{\color[HTML]{000000} Emotionally involved, Degree of Involvement}} &
  {\color[HTML]{000000} } &
  {\color[HTML]{000000} \checkmark} \\
\rowcolor[HTML]{EFEFEF} 
\cite{schroeder2001collaborating} &
  Schroeder et al. Questionnaire &
  10 &
  2001 &
  \multicolumn{1}{c|}{\cellcolor[HTML]{EFEFEF}N/A} &
   &
   \\
\rowcolor[HTML]{FFFFFF} 
{\color[HTML]{000000} \cite{larsson2001actor}} &
  {\color[HTML]{000000} Swedish Viewer-User Presence} &
  {\color[HTML]{000000} 150} &
  {\color[HTML]{000000} 2001} &
  \multicolumn{1}{c|}{\cellcolor[HTML]{FFFFFF}{\color[HTML]{000000} N/A}} &
  {\color[HTML]{000000} } &
  {\color[HTML]{000000} } \\
\rowcolor[HTML]{EFEFEF} 
\cite{takatalo2002presence} &
  \begin{tabular}[c]{@{}c@{}}Experimental Virtual\\ Environment-Experience\\ Questionnaire\end{tabular} &
  124 &
  2002 &
  \multicolumn{1}{c|}{\cellcolor[HTML]{EFEFEF}\begin{tabular}[c]{@{}c@{}}Spatial, Attention, Real, Action, Exploration,\\ Skill, Challenge, Personal relevance, Control, \\ Arousal, Valence, Flow, Being there, Impressed, Pleasant,\\ Anxiety, VE distracted, Nausea, Tiredness\end{tabular}} &
   &
  \checkmark \\
\rowcolor[HTML]{FFFFFF} 
{\color[HTML]{000000} \cite{lin2002effects}} &
  {\color[HTML]{000000} $E^{2}$I Scale} &
  {\color[HTML]{000000} 9} &
  {\color[HTML]{000000} 2002} &
  \multicolumn{1}{c|}{\cellcolor[HTML]{FFFFFF}{\color[HTML]{000000} \begin{tabular}[c]{@{}c@{}}Sensory Factor, Distraction Factor, \\ Realism Factor, Control Factor\end{tabular}}} &
  {\color[HTML]{000000} } &
  {\color[HTML]{000000} } \\
\rowcolor[HTML]{EFEFEF} 
\cite{cho2003dichotomy} &
  Cho et al. Questionnaire &
  4 &
  2003 &
  \multicolumn{1}{c|}{\cellcolor[HTML]{EFEFEF}N/A} &
   &
   \\
\rowcolor[HTML]{FFFFFF} 
{\color[HTML]{000000} \cite{nowak2003effect}} &
  {\color[HTML]{000000} Nowak \& Biocca Questionnaire} &
  {\color[HTML]{000000} 29} &
  {\color[HTML]{000000} 2003} &
  \multicolumn{1}{c|}{\cellcolor[HTML]{FFFFFF}{\color[HTML]{000000} \begin{tabular}[c]{@{}c@{}}Self-reported copresence, Perceived other's copresence,\\ Telepresence scale, Social presence\end{tabular}}} &
  {\color[HTML]{000000} } &
  {\color[HTML]{000000} } \\
\rowcolor[HTML]{EFEFEF} 
\cite{sas2003presence} &
  Sas and O'Hare Questionnaire &
  34 &
  2003 &
  \multicolumn{1}{c|}{\cellcolor[HTML]{EFEFEF}Being there, Not being there, Reflective consciousness} &
   &
   \\
\rowcolor[HTML]{FFFFFF} 
{\color[HTML]{000000} \cite{bouchard2004reliability}} &
  {\color[HTML]{000000} Bouchard et al. Questionnaire} &
  {\color[HTML]{000000} 1} &
  {\color[HTML]{000000} 2004} &
  \multicolumn{1}{c|}{\cellcolor[HTML]{FFFFFF}{\color[HTML]{000000} N/A}} &
  {\color[HTML]{000000} } &
  {\color[HTML]{000000} } \\
\rowcolor[HTML]{EFEFEF} 
\cite{vorderer2004mec} &
  \begin{tabular}[c]{@{}c@{}}MEC Spatial Presence\\ Questionnaire\end{tabular} &
  16 &
  2004 &
  \multicolumn{1}{c|}{\cellcolor[HTML]{EFEFEF}Self Location, Possible Actions} &
   &
   \\
\rowcolor[HTML]{FFFFFF} 
{\color[HTML]{000000} \cite{takatalo2004experiential}} &
  {\color[HTML]{000000} \begin{tabular}[c]{@{}c@{}}Presence-Involvement-Flow\\ Framework\end{tabular}} &
  {\color[HTML]{000000} 15} &
  {\color[HTML]{000000} 2004} &
  \multicolumn{1}{c|}{\cellcolor[HTML]{FFFFFF}{\color[HTML]{000000} \begin{tabular}[c]{@{}c@{}}Physical presence, Emotional Involvement,\\ Situational Involvement, Performance Competence\end{tabular}}} &
  {\color[HTML]{000000} } &
  {\color[HTML]{000000} } \\
\rowcolor[HTML]{EFEFEF} 
\cite{lombard2009measuring} &
  Temple Presence Inventory &
  42 &
  2009 &
  \multicolumn{1}{c|}{\cellcolor[HTML]{EFEFEF}\begin{tabular}[c]{@{}c@{}}Spatial Presence, Social presence-actor,\\ Passive social presence, Active social presence,\\ Presence as engagement, Presence as social richness,\\ Presence as social realism, Presence as perceptual realism\end{tabular}} &
   &
  \checkmark \\
\rowcolor[HTML]{FFFFFF} 
{\color[HTML]{000000} \cite{chertoff2010virtual}} &
  {\color[HTML]{000000} Virtual Experience Test} &
  {\color[HTML]{000000} 24} &
  {\color[HTML]{000000} 2010} &
  \multicolumn{1}{c|}{\cellcolor[HTML]{FFFFFF}{\color[HTML]{000000} \begin{tabular}[c]{@{}c@{}}Story Telling, Haptics,\\  Sensory Content, Task Completion, Active\end{tabular}}} &
  {\color[HTML]{000000} } &
  {\color[HTML]{000000} } \\
\rowcolor[HTML]{EFEFEF} 
\cite{hartmann2015spatial} &
  \textbf{\begin{tabular}[c]{@{}c@{}}Spatial Presence\\ Experience Scale\end{tabular}} &
  8 &
  2015 &
  \multicolumn{1}{c|}{\cellcolor[HTML]{EFEFEF}Self-location (SL) and Possible action (PA)} &
   &
  \checkmark \\
\rowcolor[HTML]{FFFFFF} 
{\color[HTML]{000000} \cite{weibel2015measuring}} &
  {\color[HTML]{000000} Self-Assessment-Manikins} &
  {\color[HTML]{000000} 6} &
  {\color[HTML]{000000} 2015} &
  \multicolumn{1}{c|}{\cellcolor[HTML]{FFFFFF}{\color[HTML]{000000} \begin{tabular}[c]{@{}c@{}}Attention Allocation, Spatial Situation Model, \\ Self-location, Possible actions, High cognitive involvement,\\ suspension of disbelief\end{tabular}}} &
  {\color[HTML]{000000} } &
  {\color[HTML]{000000} \checkmark} \\
\rowcolor[HTML]{EFEFEF} 
\cite{bailey2016does} &
  Self Presence and Spatial Presence &
  5 &
  2016 &
  \multicolumn{1}{c|}{\cellcolor[HTML]{EFEFEF}N/A} &
   &
   \\
\rowcolor[HTML]{FFFFFF} 
{\color[HTML]{000000} \cite{makransky2017development}} &
  {\color[HTML]{000000} \textbf{Multimodal Presence Scale}} &
  {\color[HTML]{000000} 15} &
  {\color[HTML]{000000} 2017} &
  \multicolumn{1}{c|}{\cellcolor[HTML]{FFFFFF}{\color[HTML]{000000} \begin{tabular}[c]{@{}c@{}}Physical Realism (PR), \\ Not paying attention to real environment (NARE),\\ Sense of being in the virtual environment (SBVE), \\ Not aware of physical mediation (NAPM),\\ Sense of coexistence (SC), Human realism (HR),\\ Not aware of artificiality of social interaction (NAASI),\\ Not aware of social mediation (NASM),\\ Sense of bodily connectivity (SBC),\\ Sense of bodily extension (SBE)\end{tabular}}} &
  {\color[HTML]{000000} } &
  {\color[HTML]{000000} } \\ \hline
\rowcolor[HTML]{EFEFEF} 
\multicolumn{2}{c}{\cellcolor[HTML]{EFEFEF}\textbf{Fidelity-based Presence Scale}} &
  \textbf{11} &
   &
  \textbf{\begin{tabular}[c]{@{}c@{}}Interaction Fidelity\\ Scenario Fidelity\\ Display Fidelity\end{tabular}} &
  \checkmark &
  \checkmark
\end{tabular}
\captionsetup{justification=centering}
\caption{Summary of presence questionnaires developed in the past and how their measurement factors compare to the FPS. We also include if the presence questionnaire employed a Delphi Study or conducted an EFA in their development. Presence questionnaires in bold are questionnaires we utilized in our Delphi study.}
\Description{Summary of presence questionnaires developed in the past and how their measurement factors compare to the FPS. We also include if the presence questionnaire employed a Delphi Study or conducted an EFA in their development. Presence questionnaires in bold are questionnaires we utilized in our Delphi study.}
\label{tab:RelatedWork}
\end{table*}

% - Call out the surveys we used in the Delphi to address why they may potentially insufficient, why they may not satisify certain use cases (respectfully)
% - Describe at the end why the FPS is more generalizable and interpretable metric via the System Fidelity framework
% - Skarbez - if there is one that fits your needs, us them. 

% - short paragraph about current literature reviews/surveys
% - here in the current state, we are going to focus on presence surveys 1-5 for <reason here>

\subsection{Past Presence Measurement Methods}

% \textbf{REFER TO DRAPER'S REVIEW ON THE FOUNDATIONS OF PRESENCE}

As highlighted in our introduction, questionnaires and surveys are one of the most prominent forms of measuring the sense of presence. This is further supported by the sheer number of efforts led to create presence surveys. In 2004, Van Baren \cite{van2004measuring} condcuted a systematic review and compiled a list of 28 surveys. In 2017, Skarbez et al. \cite{skarbez2017survey} compared and contrasted 14 presence/telepresence surveys. In 2021, Souza et al. conducted a similar review and curated a list of 29 presence surveys \cite{souza2021measuring}. However, in recent literature, it is apparent that there still exists some discourse on which questions are the most important to ask when measuring presence. Particularly, in 2024, Tran et al. \cite{tran2024survey} conducted a literature review of presence studies and found that 201 of their 320 retrieved articles leveraged previously created questionnaires with customizations applied to each survey. The sheer number of surveys reviewed plus the added concern of research performed with customized presence surveys showcases that there is no single presence survey or series of questions that address the broader needs of the research community. 

Given these concerns, we conducted a Delphi study in which we solicited a panel of expert researchers in presence. For our Delphi study, we asked our experts to review the entire contents of the Witmer-Singer Presence Questionnaire (WS-PQ) \cite{witmer1998measuring}, Igroup Presence Questionnaire (IPQ) \cite{schubert2001experience}, Slater-Usoh-Steed (SUS) presence questionnaire \cite{usoh2000using}, Multi-modal Presence Scale (MPS) \cite{makransky2017development}, and Spatial Presence Experience Scale (SPES) \cite{hartmann2015spatial}. These surveys were chosen for our Delphi study as we believe each of these surveys encompass all of the aspects of the system fidelity framework. We choose to focus on spatial presence through our work, which these surveys aim to measure. Co-presence (the sense of being with others \cite{biocca2001networked}), Social presence (the awareness of co-presence and ability to engage with others \cite{biocca2001networked}), and Self-presence (the effect of embodiment in a virtual environment \cite{biocca1997cyborg}) are other types of presence that have been investigated throughout the literature as highlighted in Table \ref{tab:RelatedWork}. With system fidelity as our guiding framework, we focus on spatial presence as it pertains to the events and interactions an individual experiences in a virtual environment (VE).

% based on the criteria that they are all highly cited works (200+ citations), are predominantly applied to VR contexts, and reflect potential nuance in phrasing and survey validation.

% For our Delphi study, we have selected to include the Witmer-Singer Presence Questionnaire (WS-PQ) \cite{witmer1998measuring}, Igroup Presence Questionnaire (IPQ) \cite{schubert2001experience}, Slater-Usoh-Steed (SUS) presence questionnaire \cite{usoh2000using}, Multi-modal Presence Scale (MPS) \cite{makransky2017development}, and Spatial Presence Experience Scale (SPES) \cite{hartmann2015spatial}. These surveys were chosen based on having garnered over 200 citations each and are widely used presence measurement tools for Virtual Reality (VR) studies.

\textit{Witmer-Singer Presence Questionnaire (WS-PQ):} The WS-PQ presence measurement tool is the longest measurement tool that we considered with 32 items. The WS-PQ loads each item into four core factors: control factors (CF), sensory factors (SF), distraction factors (DF), and realism factors (RF) \cite{witmer1998measuring}. Subsequently, items in the WS-PQ can also load into these sub-factors: involvement/control (INV/C), natural (NAT), auditory (AUD), haptic (HAPTC), resolution (RES), and interface quality (IFQUAL) \cite{witmer1998measuring}. With WS-PQ having the largest inventory in our selected surveys and a wide breadth of loading factors, the questionnaire is aimed to account for a wide range of technological and psychological facets. However, not all virtual experiences employ, for example, auditory or haptic features, which may cause participants to respond negatively to questions explicitly regarding those factors (e.g. WS-PQ question 15 and 16).

\textit{Igroup Presence Questionnaire (IPQ):} The structure of IPQ is derived from an exploratory factor analysis (EFA) and a confirmatory factor analysis (CFA). Items in IPQ's inventory include questions created by the original researchers and also questions from previous surveys such as the WS-PQ \cite{witmer1998measuring}. However, IPQ incorporates differing factor loads onto each of its items. Presence (PRES), spatial presence (SP), involvement (INV), and realness (REAL) are the primary factors a question can load into \cite{schubert2001experience}. An interesting note is that for an item to contribute as a PRES factor, it also needs to collectively load into SP, INV, and REAL \cite{schubert2001experience}. A limitation of IPQ is that it does not explicitly highlight technological factors that compose the virtual experience and impact users' sense of presence. The IPQ focuses on the disconnect users experience when immersed into a virtual environment. Subsequently, IPQ prompts users to evaluate the perceived realness of the virtual environment and their level of disconnection from the real world.

\textit{Slater-Usoh-Steed (SUS):} The SUS presence questionnaire is one of the most recognized presence measurement tools. Items in this survey were developed through as series of studies conducted by Usoh et. al \cite{usoh2000using}. This survey was created on the premise that a presence measurement tool should be capable of discriminating between real-world and virtual experiences \cite{usoh2000using}. While SUS is a widely utilized presence questionnaire, it does not employ a factor structure, unlike the other questionnaires we consider. The benefit of employing a factor stucture is that it enables researchers to draw conclusions regarding the factors as well as the overall sense of presence.

\textit{Multimodal Presence Scale (MPS):} The MPS was developed with an emphasis on Lee's \cite{lee2004presence} theoretical framework of presence \cite{makransky2019adding}. This premise led to the construction of a questionnaire that aims to address three types of presence: Physical presence, social presence, and self-presence \cite{makransky2017development}. As noted in Table \ref{tab:RelatedWork}, the MPS utilizes a wide range of factors to ensure it covers the three aforementioned presence types. As we have seen with IPQ, we witness the same utilization of CFA to compose the final items in the survey. The MPS excels in combining spatial presence, co-presence, and embodiment into a singular survey. However, it is difficult to conclude which specific components of the virtual experience contributed to the users' senses of physical, social, and self presence.

\textit{Spatial Presence Experience Scale (SPES):} The SPES survey is based on the spatial presence model proposed by Wirth et al. \cite{hartmann2015spatial, wirth2007process}. Wirth et al.'s \cite{wirth2007process} model states when users are presented a virtual or media stimulus, they are expected to exhibit a certain level of focus and subsequently develop a perception of possible actions they can perform. With this grounding, the SPES was developed through the composition of questionnaire items and performing an EFA and a CFA to include items in the finalized survey. Similarly to IPQ, the SPES focuses on the individual experience and perceptions of a virtual environment. As with previously discussed surveys, it is difficult to identify the technological components that allowed users to experience high levels of self location and personal agency. 

As we have seen, most of the aforementioned presence surveys are grounded within psychological or behavioral constructs \cite{skarbez2017survey, souza2021measuring}. While this can yield productive insights, the results are not particularly easy to translate to future design decisions. Given this gap in the literature, we specifically aimed to design a survey that would yield insights to inform key design decisions. To that end, our work introduces the FPS, a survey grounded in the system fidelity framework. The grounding in system fidelity affords us the ability to report and discuss the components of a VR system that affected sense of presence. This key feature aids in improving result interpretation and comparisons between virtual experiences which, in turn, can yield broader discussions regarding VR system design.

\subsection{The Effects of Fidelity on Presence}
In our growing understanding of presence, we note that there is a growing body of literature that highlights the idea that increased fidelity leads to higher levels of presence. McMahan et al. \cite{mcmahan2012evaluating} conducted a study evaluating display and interaction fidelity. In their study, high display fidelity and high interaction fidelity led to significantly higher presence scores. In another study, Shafer et al. \cite{shafer2019factors} conducted a study evaluating off-the-shelf VR experiences and the games' effects on cybersickness and sense of presence. They presented an initial model showcasing how realism and interactivity affect spatial presence. Adkins et al. \cite{adkins2021evaluating} conducted a study comparing grasping techniques between a tracked glove and a standard VR controller. In their work, a key outcome was that grabbing objects with the glove led to significantly higher presence scores than the controller. As we can see across multiple studies, there is a strong implication for higher fidelity environments and experiences may yield higher levels of presence.

In addition to the aforementioned studies, many of the previous questionnaires we reviewed, see Table \ref{tab:RelatedWork}, also contain components that allude to the idea that level of fidelity influences sense of presence. For instance, the WSPQ \cite{witmer1998measuring} employs auditory and haptic as factors that impact sense of presence. The MPS \cite{makransky2017development} notes physical realism as a factor that influences presence. Across previous work and questionnaire development, we consider the case that higher fidelity representations of virtual experiences can positively impact sense of presence. Therefore, the FPS explicitly follows the System Fidelity framework and is validated on its core components.

\begin{figure*}[]
    \centering
    \includegraphics[width=0.70\linewidth]{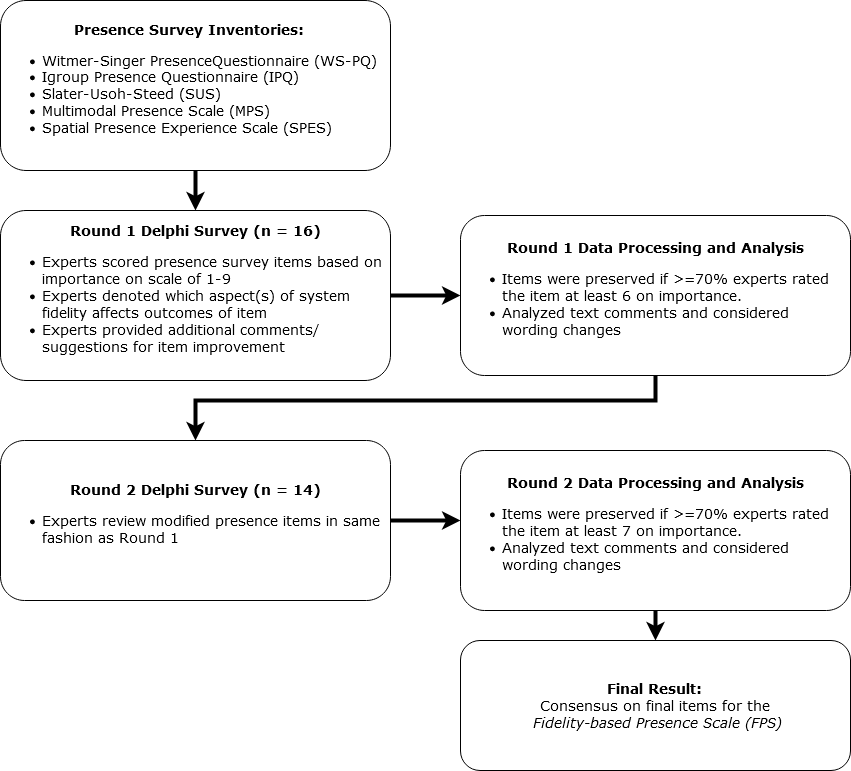}
    \caption{Summary of our Delphi process to create the Fidelity-based Presence Scale.}
    \Description{Summary of our Delphi process to create the Fidelity-based Presence Scale. It is a workflow diagram consisting of 6 bubbles, each detailing a step in our Delphi study process.}
    \label{fig:DelphiMethod}
\end{figure*}

% \subsection {Immersion}

\subsection{The Delphi Method} \label{delphiRelated}
The Delphi method is a study technique utilized to obtain consensus of opinion through administering a series of questionnaires to experts in a given area of research \cite{catete2017application, dalkey1963experimental, hsu2007delphi, keeney2006consulting}. While the definition of an ``expert'' can be debated, Adler \cite{adler1996gazing} defines an expert as an individual that has both extensive knowledge and practical experience with investigating the topic. By design, the Delphi method places a high emphasis on anonymous and iterative review wherein experts can provide their insight and feedback without potential peer pressure or conforming to a dominant view \cite{asch1956studies, milgram1963behavioral, pandor2019delphi, rowe1999delphi}. 

Previous research highlights the fact that there is no single predominant presence measurement tool or presence measurement theory \cite{skarbez2017survey, souza2021measuring, lee2004presence}. This posit is further supported by the number of presence questionnaires that have been developed over the past 40 years. By employing the Delphi Method in its development, the FPS represents a significant step forward in presence measurement as it is, to the best of our knowledge, the first consensus-driven presence measurement tool.

\section {Delphi Study Methods}
As described in section \ref{delphiRelated}, the Delphi method is a study approach designed to gain the insight and feedback of experts from a given field to form consensus on a specific topic \cite{hsu2007delphi, keeney2006consulting}. Our Delphi study methodology is adapted from Pandor et al.'s \cite{pandor2019delphi} Delphi study on Rapid Reviews. In our Delphi study, our participants completed two rounds of reviewing presence questionnaire items from WS-PQ, IPQ, SUS, MPS, and SPES. Between each round, we modified the questionnaire items based on our expert participants' feedback which were reviewed by our experts in a subsequent round of the Delphi study. Once our participants provided minimal or no feedback across the questionnaire items, we concluded our Delphi study. Figure \ref{fig:DelphiMethod} summarizes our Delphi process.

\subsection{Participant Recruitment}
\begin{table}[b]
\centering
\scriptsize
\normalsize
\begin{tabular}{lcc}
\hline
\textbf{}                 & \multicolumn{2}{c}{\textbf{Count}}                               \\ \cline{2-3} 
\textbf{Participant Characteristics} &
  \textbf{\begin{tabular}[c]{@{}c@{}}Round 1\\ (N = 16)\end{tabular}} &
  \textbf{\begin{tabular}[c]{@{}c@{}}Round 2\\ (N = 14)\end{tabular}} \\ \hline
\rowcolor[HTML]{C7C7C7} 
Gender                    &           & \multicolumn{1}{l}{\cellcolor[HTML]{C7C7C7}}         \\
\hspace{0.5 cm} Male                      & 12 (75\%) & \multicolumn{1}{c}{10 (71\%)}                        \\
\rowcolor[HTML]{C7C7C7} 
\hspace{0.5 cm} Female                    & 4 (25\%)  & \multicolumn{1}{c}{\cellcolor[HTML]{C7C7C7}4 (29\%)} \\
Country of employment     &           &                                                      \\
\rowcolor[HTML]{C7C7C7} 
\hspace{0.5 cm} Australia                 & 1 (6\%)   & 1 (7\%)                                              \\
\hspace{0.5 cm} France                    & 1 (6\%)   & 1 (7\%)                                              \\
\rowcolor[HTML]{C7C7C7} 
\hspace{0.5 cm} Germany                   & 2 (13\%)  & 2 (14\%)                                             \\
\hspace{0.5 cm} Italy                     & 2 (13\%)  & 2 (14\%)                                             \\
\rowcolor[HTML]{C7C7C7} 
\hspace{0.5 cm} Portugal                  & 2 (13\%)  & 1 (7\%)                                              \\
\hspace{0.5 cm} United Kingdom            & 1 (6\%)   & 1 (7\%)                                              \\
\rowcolor[HTML]{C7C7C7} 
\hspace{0.5 cm} USA                       & 7 (44\%)  & 6 (43\%)                                             \\     
\end{tabular}
\captionsetup{justification=centering}
\caption {A breakdown of our participant demographics of gender and global location of employment.}
\Description{A breakdown of our participant demographics of gender and global location of employment.}
\label{tab:ParticipantChar}
\end{table}

The target participants for our Delphi study were expert presence researchers. Our expert criterion included researchers that have earned their Ph.D., have a thorough history of conducting research in VR contexts, and published three or more articles that employed a standardized presence questionnaire (e.g., SUS or WS-PQ) in a VR study. This follows Adler et al.'s \cite{adler1996gazing} definition of an expert as our expert researchers had extensive knowledge along with practical experience of applying these questionnaires. We compiled an initial list of candidate experts through systematically searching electronic databases (e.g., ACM Digital Library, IEEExplore, Web of Science, Google Scholar). For databases that allowed for enhanced search parameters, we used the following query to narrow down our searches as well: \textit{TI = (presence AND virtual) OR AB = (presence AND virtual) OR AK = (presence AND virtual)}. We then gathered our expert candidates' contact information through publicly available sources (e.g., university websites) and sent personalized emails containing the definition of a Delphi study, the purpose of our study, and a Qualtrics survey link to participate in Round 1 of our Delphi study. Table \ref{tab:ParticipantChar} shows the demographics of our expert participants.

In previous Delphi studies, the appropriate number of participants varied greatly based on the domain being investigated. Broader topic Delphi studies can result in participant pools from 15 to 60 participants \cite{hasson2000research, hsu2007delphi, choe2023assessing}. Conversely, Delphi studies in more specific domains of expertise may contain 5 to 15 participants \cite{hsu2007delphi, choe2023assessing}. To ensure a response rate of at least 15 participants, given previous work, we aimed to invite at least 60 potential researchers to participate in our Delphi study. All participants were allowed to withdraw from the study at any time if needed. For all rounds of the Delphi study, participants responded through Qualtrics surveys.

\subsection{Delphi Survey Presentation}
Prior to participating in each round of the Delphi Survey, participants were presented with an introductory page that outlined the survey's purpose, listed the questionnaires included in the survey, and provided definitions for system fidelity, interaction fidelity, scenario fidelity, and display fidelity. Additionally, participants received instructions detailing the information they would encounter for each presence questionnaire item. This included guidance on providing importance ratings, specifying aspects of system fidelity that would affect the response to the item, and providing additional feedback or suggestions to improve a given item.

\subsection{Round 1} \label{round1}
\begin{figure*}[]
    \centering
    \captionsetup{justification=centering}
    \includegraphics[width=\textwidth]{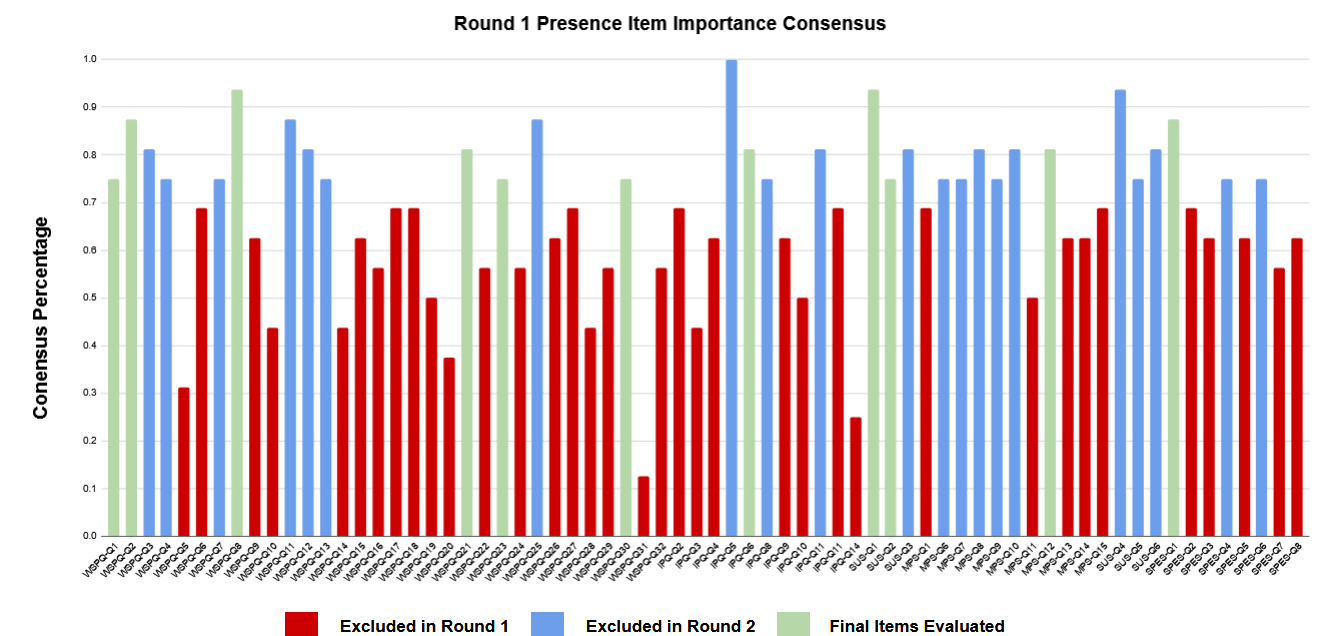}
    \caption{Round 1 summary of presence questionnaire items that met the inclusion criteria to be preserved in Round 2 of our Delphi study. Red indicates that those items were removed after Round 1 analysis. Blue indicates that those items were preserved after Round 1 and removed after Round 2. Green indicates the final items that comprise the FPS after the Delphi study.}
    \Description{Round 1 summary of presence questionnaire items that met the inclusion criteria to be preserved in Round 2 of our Delphi study. Red indicates that those items were removed after Round 1 analysis. Blue indicates that those items were preserved to be reviewed in Round 2. Green indicates the final items that comprise the FPS after the Delphi study.}
    \label{fig:FPSscores}
\end{figure*}

The Round 1 survey included the entire inventories of WS-PQ, IPQ, SUS, MPS, and SPES \cite{usoh2000using, schubert2001experience, witmer1998measuring, hartmann2015spatial, makransky2017development}. To address item duplication (e.g., IPQ Question 12 draws from WS-PQ question 7), only one instance of each duplicated item was included, resulting in a total inventory of 68 questionnaire items for review. These questionnaires were chosen on the basis that they all are highly cited surveys and we believe they encompass the aspects of the system fidelity framework.

Participants were presented with each questionnaire item in its original form, including the rating scale and anchors. Based on Pandor et al.'s \cite{pandor2019delphi} importance rating, we incorporated a Likert scale ranging from 1 (not important) to 9 (critically important) where participants rated importance of each questionnaire item. Our participants were also required to specify which component(s) of system fidelity (interaction fidelity, scenario fidelity, display fidelity) would affect the response to the item. If needed, participants were also able to offer additional suggestions to improve the clarity or grammar of the item in the form of a free-response text box. Each presence questionnaire item was accompanied with the definitions of interaction fidelity, scenario fidelity, and display fidelity. This served as a reference in case participants needed a reminder of the definitions during the survey. 

After evaluating all 68 presence items, participants were prompted to provide further suggestions or propose additional survey items. They were also asked to indicate the number of points on a Likert scale each item should be rated on and whether items should be presented as questions or statements.

With the conclusion of Round 1, we conducted an analysis of the importance ratings for each item. This analysis was performed to determine which questions would be included in Round 2 of the survey. For items that were included in the second round, we further analyzed our participants' free-text responses and applied necessary changes. We evaluated our participants' comments and re-worded the item to address our participants' concerns while maintaining the core themes and elements of the original questionnaire item. For example, SPES question 1 was initially phrased as ``I felt like I was actually there in the environment of the presentation.`` Our expert reviewers noted the term `presentation` was confusing, which led us to modify the item to be phrased as ``I felt like I was actually there in the virtual world.`` Once Round 1 concluded, participants were awarded \$50 USD in the form of an Amazon e-gift card.

\subsection{Round 2}
\label{Round2}
Given our applied changes and included items from Round 1, we believed it was necessary to conduct an additional round where our experts evaluated the presentation of the new, reduced questionnaire inventory. All the participants who completed Round 1 of the Delphi survey were invited to participate in Round 2. Along with a new Qualtrics survey link containing the revised questionnaire items, participants were provided a results package containing the overall expert panel consensus distribution for each item and their respective importance ratings as seen in Figure \ref{fig:FPSscores}.

In the Round 2 survey, experts were once again tasked with rating the importance of each survey item, specifying which component(s) of system fidelity would affect the response, and providing any additional suggestions for clarity or improvement. Each item was presented in the same form and fashion as in Round 1.

Following the conclusion of Round 2, we performed the same analyses on both our participants' perceived importance ratings and free-response comments. Once our inclusion criteria was applied, we found the final set of survey items to be sufficient as there were minimal to no comments warranting further review. Therefore, we concluded our Delphi study after Round 2. Participants were awarded \$50 USD through an Amazon e-gift card for completion of Round 2. 

\subsection{Inclusion Criteria across Round 1 and Round 2} \label{delphiCriteria}
Initially, we applied the same inclusion criteria as Pandor et al.'s \cite{pandor2019delphi} Delphi study where 70\% of experts needed to rate a given item 7 or higher in order for that item to be included in a subsequent round. However, in review of our Round 1 results, we found that criteria to drastically drop the questionnaire item count from 68 to 6. While a survey comprising of 6 items could be considered sufficient, we believed that there were numerous items that should be reconsidered. Jorm \cite{jorm2015using} notes that in cases where there are a high number of items for experts to evaluate, it is common for items to be advocated for reconsideration once during the Delphi study, such as in Yap et al.'s work \cite{yap2014parenting}. Therefore, for Round 1, we included questionnaire items that received a rating of 6 or higher from 70\% or more of our experts. In Round 2, we maintained the original inclusion criteria of at least 70\% of experts needing to rate an item 7 or higher as our experts' did not present new concerns/changes to any of the items in the survey.

Additionally, our criteria for individual item changes followed the workflow highlighted in section \ref{round1}. For each item, we considered comments and concerns that were raised by \textit{at least two} of our experts. This allowed for strict changes to be applied across items if needed. We also categorized changes as either being a global change or a local change. For global changes, these were changes that were expressed across multiple items that our experts believed should be modified. For example, SPES \#1 was phrased as `` I felt like I was actually there in the environment of the presentation.'' Our experts first noted their concern with the phrasing ``environment of the presentation'' which was found across multiple items. Given that this concern of phrasing was raised across multiple items, we applied a global change in which we ensured all items would be phrased with "virtual world" as seen in Table \ref{tab:FPSImportanceTable}. A local change is one that we applied specifically to an item. For example, in our Round 1 review, the original phrasing of WSPQ \#1 was ``How much were you able to control events?''. Four of our experts noted the potential confusion of the phrasing ``control events'', which led us to change the phrasing of this item to "I felt I could control my actions in the virtual world". Our comment/concern count criteria is also the reason we resulted in two rounds for our Delphi study. We conducted additional rounds if two or more experts raised concerns for a given item. In our case, our experts did not raise concerns after reviewing the changed items in round 2, which led to us finalizing our Delphi study.

\subsection{Delphi Study Results}
In summary, our Delphi study is comprised of two rounds of presence questionnaire item review. Data for Round 1 was collected over the span of 2 months, and data for Round 2 was also collected over the course of 2 months. In this section, we detail the key results and outcomes of our Delphi study.

\begin{table*}[h]
\scriptsize
\centering
\begin{tabular}{ccc|cc}
\rowcolor[HTML]{FFFFFF} 
\textbf{\begin{tabular}[c]{@{}c@{}}Item\\ Origin\end{tabular}} &
  \textbf{\begin{tabular}[c]{@{}c@{}}Original \\ Item\end{tabular}} &
  \textbf{\begin{tabular}[c]{@{}c@{}}FPS \\ Item\end{tabular}} &
  \textbf{\begin{tabular}[c]{@{}c@{}}Round 1 \\ Consensus\end{tabular}} &
  \textbf{\begin{tabular}[c]{@{}c@{}}Round 2 \\ Consensus\end{tabular}} \\ \hline
\rowcolor[HTML]{EFEFEF} 
SUS-Q1 &
  \textbf{\begin{tabular}[c]{@{}c@{}}Please rate your sense of being in the environment, on the \\ following  scale from 1 to 7,  where 7 represents \\ your normal experience of being in a place\end{tabular}} &
  \textbf{\begin{tabular}[c]{@{}c@{}}My sense of being in the virtual world\\ was like being in a real place\end{tabular}} &
  15/16 (93\%) &
  11/14 (78\%) \\
\rowcolor[HTML]{FFFFFF} 
SUS-Q2 &
  \textbf{\begin{tabular}[c]{@{}c@{}}To what extent were there times during the experience \\ when the virtual environment was reality for you?\end{tabular}} &
  \textbf{\begin{tabular}[c]{@{}c@{}}During the experience,  I felt the \\ virtual world was reality for me.\end{tabular}} &
  12/16 (75\%) &
  10/14 (71\%) \\
\rowcolor[HTML]{EFEFEF} 
SPES-Q1 &
  \textbf{\begin{tabular}[c]{@{}c@{}}I felt like I was actually there in the\\ environment of the presentation\end{tabular}} &
  \textbf{\begin{tabular}[c]{@{}c@{}}I felt like I was actually \\ there in the virtual world.\end{tabular}} &
  14/16 (87\%) &
  10/14 (71\%) \\
\rowcolor[HTML]{FFFFFF} 
IPQ-Q6 &
  \textbf{I felt present in the virtual space} &
  \textbf{\begin{tabular}[c]{@{}c@{}}I felt present in the \\ virtual world.\end{tabular}} &
  13/16 (81\%) &
  11/14 (78\%) \\
\rowcolor[HTML]{EFEFEF} 
WSPQ-Q1 &
  \textbf{\begin{tabular}[c]{@{}c@{}}How much were you able to control\\ events?\end{tabular}} &
  \textbf{\begin{tabular}[c]{@{}c@{}}I felt I could control my actions \\ within the virtual world.\end{tabular}} &
  12/16 (75\%) &
  12/14 (86\%) \\
\rowcolor[HTML]{FFFFFF} 
WSPQ-Q2 &
  \textbf{\begin{tabular}[c]{@{}c@{}}How responsive was the environment\\ to the actions that you initiated (or performed)?\end{tabular}} &
  \textbf{\begin{tabular}[c]{@{}c@{}}I felt the virtual world was \\ responsive to my actions.\end{tabular}} &
  14/16 (87\%) &
  12/14 (86\%) \\
\rowcolor[HTML]{EFEFEF} 
WSPQ-Q8 &
  \textbf{\begin{tabular}[c]{@{}c@{}}How aware were you of events\\ occurring in the real world around you?\end{tabular}} &
  \textbf{\begin{tabular}[c]{@{}c@{}}I was not aware of events occurring \\ in the real world around me.\end{tabular}} &
  15/16 (93\%) &
  10/14 (71\%) \\
\rowcolor[HTML]{FFFFFF} 
WSPQ-Q21 &
  \textbf{\begin{tabular}[c]{@{}c@{}}How well could you move or manipulate\\ objects in the virtual environment?\end{tabular}} &
  \textbf{\begin{tabular}[c]{@{}c@{}}I felt I could move or manipulate objects \\ easily in the virtual world.\end{tabular}} &
  13/16 (81\%) &
  10/14 (71\%) \\
\rowcolor[HTML]{EFEFEF} 
WSPQ-Q23 &
  \textbf{\begin{tabular}[c]{@{}c@{}}How involved were you in the virtual\\ environment experience?\end{tabular}} &
  \textbf{\begin{tabular}[c]{@{}c@{}}I felt involved in the \\ virtual world experience.\end{tabular}} &
  12/16 (75\%) &
  \cellcolor[HTML]{EFEFEF}10/14 (71\%) \\
\rowcolor[HTML]{FFFFFF} 
WSPQ-Q30 &
  \textbf{\begin{tabular}[c]{@{}c@{}}How well could you concentrate on the assigned tasks \\ or required activities rather than on the mechanisms \\ used to perform those tasks or activities?\end{tabular}} &
  \textbf{\begin{tabular}[c]{@{}c@{}}I could concentrate on the virtual \\ activities rather than the controls\\ to perform them.\end{tabular}} &
  12/16 (75\%) &
  \cellcolor[HTML]{FFFFFF}10/14 (71\%) \\
\rowcolor[HTML]{EFEFEF} 
MPS-Q12 &
  \textbf{\begin{tabular}[c]{@{}c@{}}When something happened to my\\ virtual embodiment, it felt like it was\\ happening to my real body.\end{tabular}} &
  \textbf{\begin{tabular}[c]{@{}c@{}}When something happened to my \\ virtual body, I felt it happened to my \\ real body.\end{tabular}} &
  13/16 (81\%) &
  \cellcolor[HTML]{EFEFEF}10/14 (71\%)
\end{tabular}
\captionsetup{justification=centering}
\caption{Summary of expert importance consensus across Round 1 and Round 2 along with the percentage of experts that identified a component of system fidelity that would affect the outcome of a question after Round 2.}
\Description{Summary of expert importance consensus across Round 1 and Round 2 along with the percentage of experts that identified a component of system fidelity that would affect the outcome of a question after Round 2.}
\label{tab:FPSImportanceTable}
\end{table*}

\subsubsection{Round 1 Consensus}
Out of the 68 items presented, 32 items (47\%) garnered a consensus of 70\% or higher in terms of importance. Participants' comments emphasized maintaining consistency in wording and Likert scale rating range for each item. As detailed in section \ref{delphiCriteria}, we conducted a thorough analysis of each questionnaire item individually, and the Delphi survey was revised to present each of the 32 items as statements and to utilize a 7-point Likert scale for rating based on our experts' feedback.

\subsubsection{Round 2 Consensus}
In Round 2, 11 items (28\%) out of the 32 items achieved a consensus rating of 70\% or higher from our experts. Overall, our participants expressed satisfaction towards the inventory of the survey and no additional suggestions were made for improving the survey within the 11 items. Following a thorough review of our experts' comments and potential suggestions, Round 2 concluded the Delphi study as there were no outstanding suggestions or views that warranted additional rounds. Furthermore, we also wanted to prevent our participants from exhibiting survey fatigue from the iterative process \cite{keeney2006consulting}. The resulting version of our presence survey, including the importance rating consensus across both rounds, each item's origin, and which components of system fidelity affect the item, is presented in Table \ref{tab:FPSImportanceTable}. 

\section{Factor Analysis Validation Study} \label {EFAStudy}
After the conclusion of our Delphi study, the FPS needed to be validated through a user study. Given FPS' theoretical basis in system fidelity, we conducted a $2\times2\times2$ ($Interaction Fidelity \times Scenario Fidelity \times Display Fidelity$) within-subjects VR user study. With participants responding to the FPS within each condition, we applied a Latin square design to ensure conditions were counterbalanced. For all eight conditions, participants were immersed into a kitchen environment with the responsibility of putting dishes away.

\subsection{Independent Variables}
We investigated three independent variables in our study: interaction fidelity, scenario fidelity, and display fidelity. For each of the system fidelity components, we implemented a low and high fidelity variant to evaluate. Each of our implementations were not intended to comprehensively represent all the aspects of each fidelity construct, but rather focus on specific elements that directly influence user experience in our task context. We further detail which aspects of each fidelity component we include in our study design.
% \ReviewerThree{We focused on specific sub-components of each fidelity construct that were most relevant to our research objectives. For Scenario Fidelity, we concentrated on Physical Coherence by manipulating the effects of gravity on objects, while acknowledging that other sub-components like Behavioral Coherence and Attribute Coherence were not included. For display fidelity, we varied both visual and auditory fidelities by adjusting environmental textures, object models, and incorporating sounds and background music. We did not include tactile fidelity in our implementations. Each of our implementations was not intended to comprehensively represent all aspects of each fidelity construct, but rather to focus on specific elements that directly influence user experience in our task context.}

\subsubsection{Interaction Fidelity}
Since the main task was to grab and put dishes away, we implemented two techniques to represent high interaction fidelity and low interaction fidelity. For high interaction fidelity, we used direct manipulation with virtual hands \cite{poupyrev1998egocentric}. This approach included predefined hand gestures for direct manipulation when grabbing objects, allowing for a more intuitive interaction style as seen in Sub-figure A from Figure \ref{fig:TeaserFigure}. Conversely, we adapted Bowman et al.'s \cite{bowman1997evaluation} ray cast ``fishing reel'' metaphor technique as our low interaction fidelity condition. A ray cast based technique was utilized as it is noted in previous work \cite{bowman1995user, bowman1997evaluation, poupyrev1996go} that ray cast interaction techniques increase the difficulty of rotating the object in place. Sub-figures A and D from Figure \ref{fig:TeaserFigure} shows the representation of both the interaction technique levels. Between our two selection techniques, direct manipulation with the VR controllers provides a higher level of biomechanical symmetry (i.e. degree with which real-world body movements are reproduced with an interaction \cite{mcmahan2016affect, ragan2015effects}) and higher level of control symmetry (i.e. the degree with which control in a task is provided by an interaction \cite{mcmahan2016affect, ragan2015effects}) than the ray-cast based technique.

\subsubsection{Scenario Fidelity}
We implemented two techniques to represent high scenario fidelity and low scenario fidelity. In high scenario fidelity conditions, gravity would always be enabled and dishes would appropriately drop when let go. In the low scenario fidelity conditions, when the participants would let go of a dish, there was a 50\% chance that the gravity would be disabled for that single dish. This would lead to instances where some plates would be floating around and some plates would be affected by gravity. Sub-figures B and E from Figure \ref{fig:TeaserFigure} shows the two variations of scenario fidelity. With our scenario fidelity conditions, we were able to vary the physical coherence (i.e. how consistent the physics of the virtual environment are to the real world \cite{skarbez2017psychophysical, mcmahan2024virtual}) and attribute coherence (i.e. how consistent attributes of virtual objects are to their real world counterpart \cite{mcmahan2024virtual}). However, due to our study design not including virtual agents, we cannot evaluate behavioral coherence as it is contingent on the quality and consistency of virtual agent behaviors \cite{mcmahan2024virtual}.

\subsubsection{Display Fidelity}
Throughout the study, participants were immersed into two variants of the kitchen environment. We explored display fidelity by varying visual fidelity (i.e. the degree with which realistic visuals are reproduced in a virtual environment \cite{mcmahan2024virtual, ragan2015effects}) and auditory fidelity (i.e. the degree with which realistic audio stimuli are reproduced in a virtual environment \cite{mcmahan2024virtual, ragan2015effects}). In the low display fidelity condition, the textures of the environment were minimal and there was no sound in the environment. Additionally, the dishes were primarily primitive Unity objects (i.e. thin cylinders). In the high display fidelity condition, realistic textures were present, the dishes made noises when colliding with other objects, there was royalty-free background music with spatial audio enabled, and the dishes were actual plate and bowl models. Sub-figures C and F from Figure \ref{fig:TeaserFigure} shows the two display fidelity environments in our study.

% \begin{figure}[]
%     \centering
%     \includegraphics[width=0.5\linewidth]{images/FPS-Environments.png}
%     \caption{The high display fidelity environment (top) and the low display fidelity environment (bottom) participants experienced during our validation user study}
%     \label{fig:studyEnvironments}
% \end{figure}

\subsection{User Study Procedure}
Participants recruited for our study were asked to review and complete a pre-screening survey through Qualtrics. In the survey, participants were asked to review an informed consent document, an eligibility document, and provide demographic information, such as age and gender. Following the conclusion of the survey, participants were then asked to schedule a day and time in which they would be able to participate in the in-person VR study.

On arrival, participants were introduced to the Meta Quest Pro, the VR system used to administer the study. This introduction included how to adjust the headset for comfort as well as the controls needed to perform the tasks in the study. We then informed participants on the overall task. We explained to participants that the study is comprised of eight trials of putting dishes away. Each trial was comprised of putting away 20 dishes in any of the cabinets within the kitchen environment. We also explained that between each trial, aspects of the virtual world may change (interaction fidelity, scenario fidelity, and display fidelity). Participants would know if all the dishes were put away once the in-VR FPS questionnaire was presented for them to respond to. The environment is based on the designs presented by Hmaiti et al. \cite{hmaiti2024visual}.

After explaining the procedure, participants were immersed into the eight trials. For each trial, participants would respond to the entire inventory of the FPS on a 1 to 7-point Likert scale. After four trials, participants were offered the opportunity to take a break from VR. After their break, participants would complete the remaining four trials of putting dishes away.

Once all eight trials were completed, participants were asked if they had any comments or questions regarding their experience. The overall time required to complete the study was approximately 60 minutes, and participants were compensated \$20 USD via an Amazon e-gift card.

\subsection{Participants}
We recruited a total of 55 participants (25 female, 30 male) from our local university. Participants were required to be 18 years of age or older, have normal or corrected-to-normal vision, and be able to hear, walk, extend both arms, use both hands, and speak and understand English. Participants with any visual, auditory, neurological, or physical disabilities were excluded. The ages of our participants ranged from 18 to 32 with a mean age of 21.62.

\subsection{Exploratory Factor Analysis}
In our validation study, 55 participants responded to the FPS under 8 unique conditions, totaling 440 complete responses to the FPS. Following our validation study, we conducted an exploratory factor analysis (EFA) to reveal further insights into the factors each item loads onto and if there are items that need to be removed. 

On the initial data collected, we constructed a correlation matrix to learn if there were items in the FPS that were highly correlated. We found no item pairings from the original FPS inventory with correlation values higher than 0.9. We then verified the sampling adequacy of our data using the Kaiser-Meyer-Olkin (KMO) measure, KMO = 0.9, which indicates that our data is suitable for factor analysis \cite{kaiser1974little, kaiser1970second}. Our Bartlett's test of sphericity ($\chi^{2} = 3270.034, p < 0.001$) was significant and indicated the presence of correlations among the FPS' survey items \cite{bartlett1950tests}.

\begin{table*}[]
\begin{tabular}{c|cccccc}
\cline{2-7}
                                & \multicolumn{3}{c|}{\cellcolor[HTML]{FFFFFF}\textbf{Delphi Results}} & \multicolumn{3}{c|}{\cellcolor[HTML]{FFFFFF}\textbf{EFA Results}} \\ \hline
\rowcolor[HTML]{FFFFFF} 
\multicolumn{1}{|c|}{\cellcolor[HTML]{FFFFFF}{\color[HTML]{000000} \textbf{Item}}} &
  \textbf{\begin{tabular}[c]{@{}c@{}}Interaction\\ Fidelity\\ Consensus\end{tabular}} &
  \textbf{\begin{tabular}[c]{@{}c@{}}Scenario\\ Fidelity\\ Consensus\end{tabular}} &
  \multicolumn{1}{c|}{\cellcolor[HTML]{FFFFFF}\textbf{\begin{tabular}[c]{@{}c@{}}Display\\ Fidelity\\ Consensus\end{tabular}}} &
  \textbf{\begin{tabular}[c]{@{}c@{}}Interaction \\ Fidelity\end{tabular}} &
  \textbf{\begin{tabular}[c]{@{}c@{}}Scenario\\ Fidelity\end{tabular}} &
  \multicolumn{1}{c|}{\cellcolor[HTML]{FFFFFF}\textbf{\begin{tabular}[c]{@{}c@{}}Display \\ Fidelity\end{tabular}}} \\ \hline
\rowcolor[HTML]{EFEFEF} 
SUS-Q1                          & 7/14 (50\%)           & 13/14 (93\%)          & 11/14 (78\%)         & \textbf{0.40}        & 0.16                 & \textbf{0.77}       \\
\rowcolor[HTML]{FFFFFF} 
{\color[HTML]{000000} SUS-Q2}   & 10/14 (71\%)          & 14/14 (100\%)         & 10/14 (71\%)         & \textbf{0.31}        & 0.18                 & \textbf{0.84}       \\
\rowcolor[HTML]{EFEFEF} 
SPES-Q1                         & 10/14 (71\%)          & 11/14 (78\%)          & 11/14 (78\%)         & 0.24                 & 0.40                 & \textbf{0.70}       \\
\rowcolor[HTML]{FFFFFF} 
{\color[HTML]{000000} IPQ-Q6}   & 12/14 (86\%)          & 10/14 (71\%)          & 11/14 (78\%)         & \textbf{0.34}        & \textbf{0.86}        & \textbf{0.36}       \\
\rowcolor[HTML]{EFEFEF} 
WSPQ-Q1                         & 14/14 (100\%)         & 4/14 (29\%)           & 1/14 (7\%)           & \textbf{0.86}        & 0.20                 & 0.23                \\
\rowcolor[HTML]{FFFFFF} 
{\color[HTML]{000000} WSPQ-Q2}  & 14/14 (100\%)         & 9/14 (64\%)           & 3/14 (21\%)          & \textbf{0.89}        & 0.19                 & 0.20                \\
\rowcolor[HTML]{EFEFEF} 
WSPQ-Q8                         & 3/14 (21\%)           & 5/14 (36\%)           & 10/14 (71\%)         & 0.15                 & 0.10                 & 0.30                \\
\rowcolor[HTML]{FFFFFF} 
{\color[HTML]{000000} WSPQ-Q21} & 12/14 (86\%)          & 7/14 (50\%)           & 2/14 (14\%)          & \textbf{0.73}        & 0.10                 & \textbf{0.35}       \\
\rowcolor[HTML]{EFEFEF} 
WSPQ-Q23                        & 9/14 (64\%)           & 11/14 (78\%)          & 6/14 (43\%)          & \textbf{0.51}        & \textbf{0.37}        & \textbf{0.37}       \\
\rowcolor[HTML]{FFFFFF} 
{\color[HTML]{000000} WSPQ-Q30} & 14/14 (100\%)         & 2/14 (14\%)           & 4/14 (29\%)          & \textbf{0.58}        & 0.19                 & 0.29                \\
\rowcolor[HTML]{EFEFEF} 
MPS-Q12                         & 8/14 (57\%)           & 5/14 (36\%)           & 11/14 (78\%)         & 0.13                 & 0.1                  & \textbf{0.61}      
\end{tabular}
\captionsetup{justification=centering}
\caption{The Structure Matrix after running EFA after performing an orthogonal rotation on the 11 questions. Values greater than |0.3| are in bold based on criteria from Hair et al. \cite{hair2009multivariate}. Alongside the matrix are the experts' rating distributions of which component of System Fidelity would affect each of the FPS items}
\Description{The Structure Matrix after running EFA after performing an orthogonal rotation on the 11 questions. Values greater than |0.3| are in bold based on criteria from Hair et al. \cite{hair2009multivariate}. Alongside the matrix are the experts' rating distributions of which component of System Fidelity would affect each of the FPS items}
\label{tab:firstEFA}
\end{table*}

To determine the appropriate number of factors to apply to our survey, we first computed the eigenvalues to learn of the number of factors Kaiser's criterion recommends. We also conducted a parallel analysis to compare the factor valuation Kaiser's criterion recommended. Our eigenvalues, based on Kaiser's criterion of 1, recommends that we maintain two factors while our parallel analysis suggested four factors. Given our experts' consensus on three factors in our Delphi Study, we conducted our initial EFA with three factors, which falls within the suggested ranges from Kaiser's criterion and the parallel analysis. For each of the structure matrices, we applied an inclusion threshold of > $|0.3|$ since our user study resulted in 440 samples of completed FPS surveys. This follows the factor analysis inclusion criteria based on sample size set by Hair et al. \cite{hair2009multivariate}.

In Table \ref{tab:firstEFA}, we present the initial structure matrix for the original 11 survey items. Upon running the EFA, we note that WSPQ-Q8  did not sufficiently load into any of the identified factors of interaction fidelity, display fidelity, and scenario fidelity as seen in Table \ref{tab:firstEFA}. Therefore, we removed WSPQ-Q8  from our final item pool and conducted an additional EFA on the remaining survey items.

In the revised data, we constructed a correlation matrix and found no item pairings to have correlation values higher than 0.9. The KMO mean sampling adequacy maintained a value of 0.9. Our Bartlett's test of sphericity (($\chi^{2} = 3178.535, p < 0.001$) remained significant as well. As with the original data, we also computed the eigenvalues and conducted a parallel analysis. Kaiser's criterion recommended 2 factors and the parallel analysis recommended 3 factors are appropriate for the revised data. Table \ref{tab:finalEFA} presents the structure matrix on the revised data along with the calculated Cronbach's alpha for each of the factors. None of our factors were found to be violation of the Cronbach's alpha threshold as all values were greater than 0.7 \cite{cronbach1951coefficient}.

\begin{table*}[]
\begin{tabular}{cccc}
\multicolumn{1}{c|}{\textbf{FPS Item}} & \textbf{Interaction Fidelity} & \textbf{Scenario Fidelity} & \textbf{Display Fidelity} \\ \hline
\rowcolor[HTML]{EFEFEF} 
\multicolumn{1}{c|}{\cellcolor[HTML]{EFEFEF}\textbf{FPS-1}} & \textbf{0.40} & 0.18          & \textbf{0.77} \\
\multicolumn{1}{c|}{\textbf{FPS-2}}                         & \textbf{0.32} & 0.20          & \textbf{0.84} \\
\rowcolor[HTML]{EFEFEF} 
\multicolumn{1}{c|}{\cellcolor[HTML]{EFEFEF}\textbf{FPS-3}} & 0.20          & \textbf{0.41} & \textbf{0.69} \\
\multicolumn{1}{c|}{\textbf{FPS-4}}                         & \textbf{0.34} & \textbf{0.87} & \textbf{0.35} \\
\rowcolor[HTML]{EFEFEF} 
\multicolumn{1}{c|}{\cellcolor[HTML]{EFEFEF}\textbf{FPS-5}} & \textbf{0.86} & 0.21          & 0.23          \\
\multicolumn{1}{c|}{\textbf{FPS-6}}                         & \textbf{0.89} & 0.20          & 0.19          \\
\rowcolor[HTML]{EFEFEF} 
\multicolumn{1}{c|}{\cellcolor[HTML]{EFEFEF}\textbf{FPS-7}} & \textbf{0.73} & 0.11          & \textbf{0.34} \\
\multicolumn{1}{c|}{\textbf{FPS-8}}                         & \textbf{0.51} & \textbf{0.38} & \textbf{0.36} \\
\rowcolor[HTML]{EFEFEF} 
\multicolumn{1}{c|}{\cellcolor[HTML]{EFEFEF}\textbf{FPS-9}} & \textbf{0.58} & 0.20          & 0.28          \\
\multicolumn{1}{c|}{\textbf{FPS-10}}                        & 0.14          & 0.12          & \textbf{0.60} \\ \hline
\rowcolor[HTML]{EFEFEF} 
\textbf{Cronbach's Alpha}                                   & \textbf{0.91} & \textbf{0.82} & \textbf{0.89}
\end{tabular}
\captionsetup{justification=centering}
\caption{The Structure Matrix after running EFA with an orthogonal rotation on the revised data. Values greater than |0.3| are in bold based on criteria from Hair et al. \cite{hair2009multivariate}. The final row reveals sufficient Cronbach's alpha values across each of the factors.}
\label{tab:finalEFA}
\end{table*}

\begin{figure}[h]
    \centering
    \includegraphics[width=0.65\linewidth]{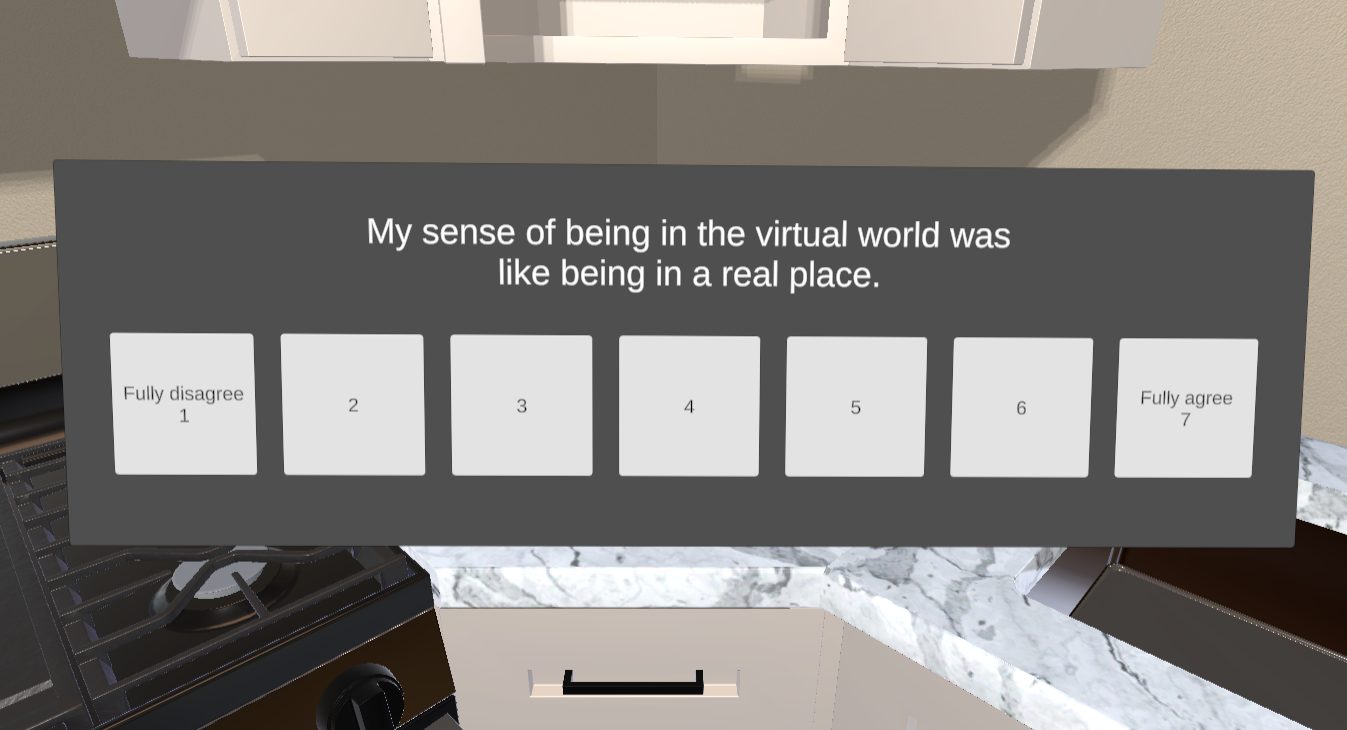}
    \captionsetup{justification=centering}
    \caption{How an item from the FPS was presented in the virtual environment. Each item is rated on a 7-point Likert scale with the anchors: "Fully disagree" and "Fully agree"}
    \Description{How an item from the FPS was presented in the virtual environment. Each item is rated on a 7-point Likert scale with the anchors: "Fully disagree" and "Fully agree"}
    \label{fig:FPSItem}
\end{figure}

\subsection{The Fidelity-based Presence Scale (FPS)}
The final version of the FPS consists of 10 items that can be utilized across VR experiments that aim to evaluate participants' sense of presence in virtual environments. Across the Delphi study and our validation study, we have shown the development process of the FPS from theory into practice. Based on feedback from our experts, the FPS should be collected utilizing a 7-point Likert scale ranging from "Fully Disagree" to "Fully Agree". Figure \ref{fig:FPSItem} shows how an item from the FPS would be employed in a virtual reality context.

\subsection{Calculating the FPS Score}
Results of data collected using the FPS will be presented across four scores: Interaction Presence Score, Scenario Presence Score, Display Presence Score, and Total Presence Score. The purpose of breaking down the results into these four scores to allow future work to provide additional insight into which features of the environment attributed to overall sense of presence.

\begin{itemize}
    \item Interction Presence Score: (FPS-1 + FPS-2 + FPS-4 + FPS-5 + FPS-6 + FPS-7 + FPS-8 + FPS-9) / 8
    \item Scenario Presence Score: (FPS-3 + FPS-4 + FPS-8) / 3
    \item Display Presence Score: (FPS-1 + FPS-2 + FPS-3 + FPS-4 + FPS-7 + FPS-8 + FPS-10) / 7
    \item Total Presence Score: ($\sum_{n=1}^{10} FPS_{n}$) / 10
\end{itemize}

\subsection{Reporting Results of the FPS} \label{resultReport}
After collecting our FPS results across our 55 participants, we calculated each of the presence scores, as highlighted in section 4.6. We found the data to violate normality, therefore we normalized our data through the Aligned Rank Transform (ART) approach \cite{wobbrock2011aligned}. For each presence score, we report the analysis of variance (ANOVA) and their corresponding pairwise t-test(s) for the main and interaction effects.

\subsection{Total Presence Score}
For our Total Presence Score, our repeated measures ANOVA revealed significant main effects for interaction fidelity $F_{1, 54} = 180.42, p < 0.001, \eta^2 = 0.77$ and scenario fidelity $F_{1, 54} = 39.26, p < 0.001, \eta^2 = 0.421$. Our analysis also revealed a significant interaction effect between interaction fidelity and display fidelity $F_{1, 54} = 7.01, p = 0.008, \eta^2 = 0.115$. Our post-hoc tests for interaction fidelity revealed that virtual hands yielded significantly higher Total Presence Scores than the ray cast interaction technique $t_{54} = 13.43, p < 0.001$. Similarly, post-hoc tests for scenario fidelity revealed that consistent gravity yielded higher total presence scores than inconsistent gravity $t_{54} = 6.27, p < 0.001$. Table \ref{tab:TOTPresScoreTable} shows the post-hoc tests for the interaction effect between interaction fidelity and display fidelity. Figure \ref{fig:MeanPresScores} also highlights the mean Total Presence Scores from our study.

\begin{table*}[t]
\begin{tabular}{cc|cc}
\textbf{Condition Pairing 1}             & \textbf{Condition Pairing 2}   & \textbf{t-score} & \textbf{Significance} \\ \hline
\rowcolor[HTML]{EFEFEF} 
\textbf{High Interaction x High Display} & High Interaction x Low Display & $t_{54} = 3.23$  & p = 0.008             \\
\textbf{High Interaction x High Display} & Low Interaction x High Display & $t_{54} = 11.62$ & p \textless 0.001 \\
\rowcolor[HTML]{EFEFEF} 
\textbf{High Interaction x High Display} & Low Interaction x Low Display  & $t_{54} = 10.73$ & p \textless 0.001 \\
\textbf{High Interaction x Low Display}  & Low Interaction x High Display & $t_{54} = 8.39$  & p \textless 0.001 \\
\rowcolor[HTML]{EFEFEF} 
\textbf{High Interaction x Low Display}  & Low Interaction x Low Display  & $t_{54} = 7.50$  & p \textless 0.001
\end{tabular}
\captionsetup{justification=centering}
\caption{Post-hoc tests highlighting the interaction effect for interaction fidelity and display fidelity in regard to the Total Presence Score. Condition pairings in bold are condition parameters that received higher Interaction Presence Scores than their corresponding pairing in the same row.}
\Description{Post-hoc tests highlighting the interaction effect for interaction fidelity and display fidelity in regard to the Total Presence Score. Condition pairings in bold are condition parameters that received higher Interaction Presence Scores than their corresponding pairing in the same row.}
\label{tab:TOTPresScoreTable}
\end{table*}

% \begin{figure}[b]
%     \centering
%     \includegraphics[width=0.6\linewidth]{images/MeanIntScores-R.png}
%     \captionsetup{justification=centering}
%     \caption{The mean Interaction Presence Scores between all low interaction fidelity and high interaction fidelity conditions.}
%     \label{fig:MeanIntScores}
% \end{figure}

\subsection{Interaction Presence Score}
For our Interaction Presence Score, we conducted an ANOVA, which found significance with regard to interaction fidelity level $F_{1, 54} = 171.11, p < 0.001, \eta^2 = 0.76$. For high interaction fidelity conditions, we had a mean of 5.28 with standard deviation of 0.995. For low interaction fidelity conditions, we had a mean of 4.20 and a standard deviation of 1.22. Figure \ref{fig:FidelityMeans} highlights the mean Interaction Presence Scores from our validation study across low interaction and high interaction fidelity conditions.

\begin{figure*}[t]
    \centering
    \includegraphics[width= 0.85\textwidth]{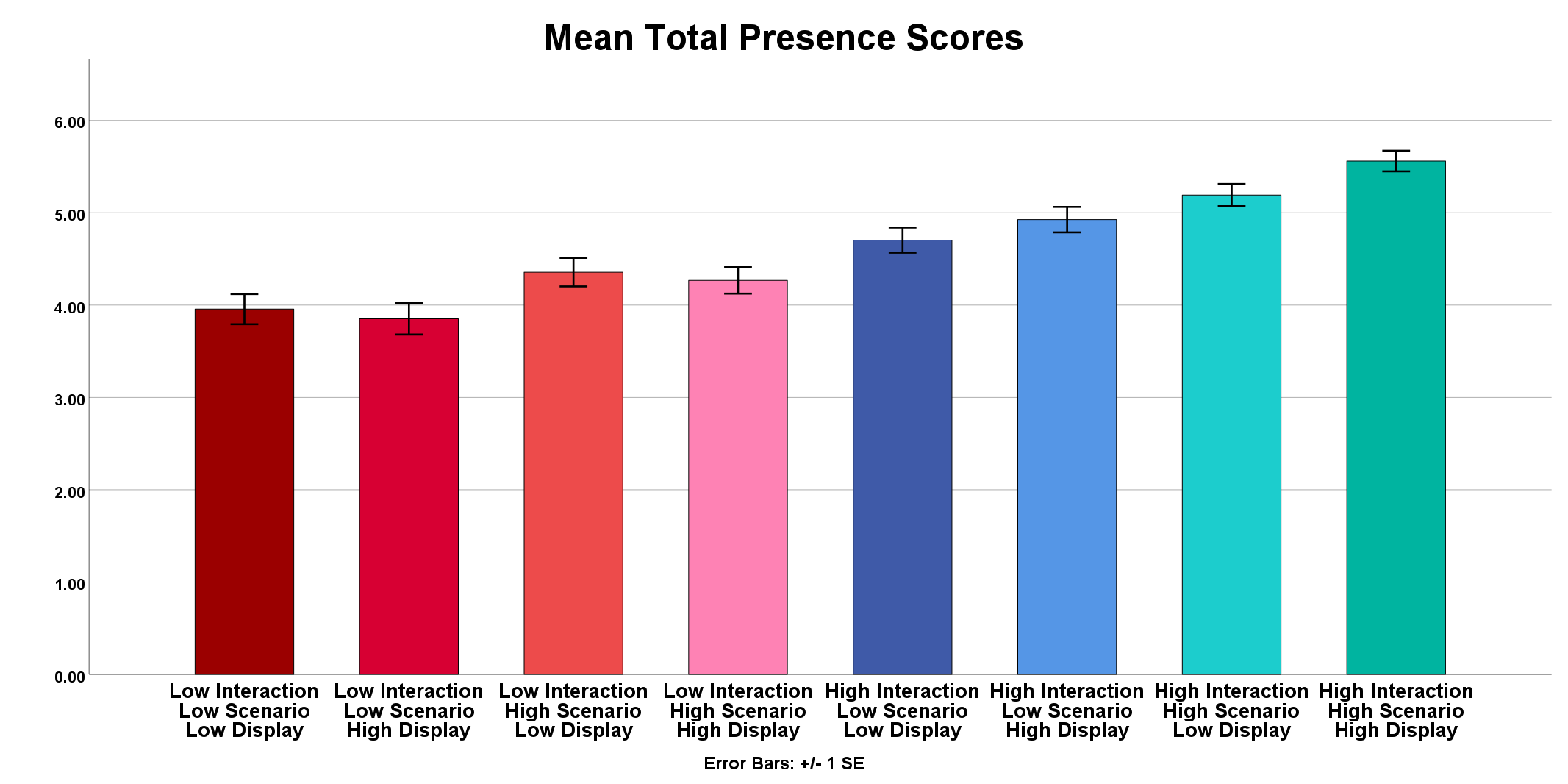}
    \captionsetup{justification=centering}
    \caption{The mean Total Presence Scores across the 8 conditions of our validation study}
    \Description{The mean Total Presence Scores across the 8 conditions of our validation study}
    \label{fig:MeanPresScores}
\end{figure*}

\begin{figure}[b]
    \centering
    \includegraphics[width=\linewidth]{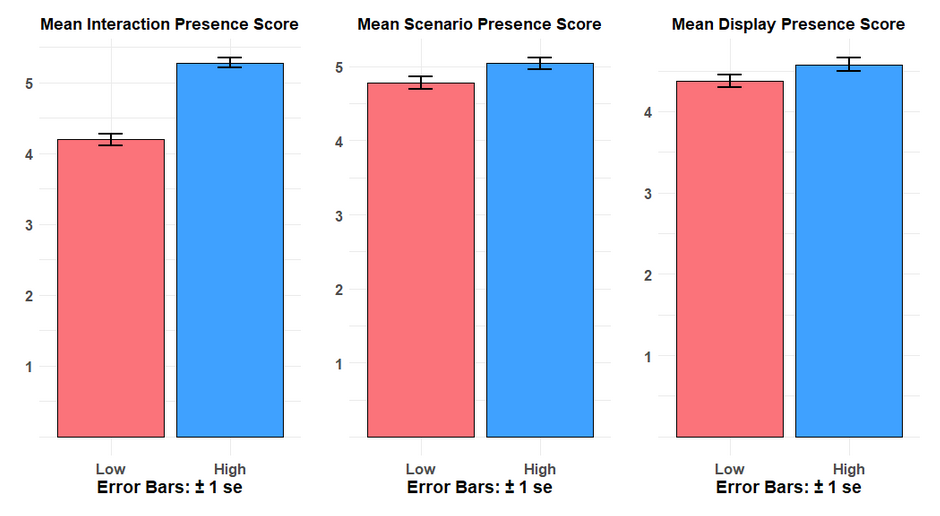}
    \captionsetup{justification=centering}
    \caption{The mean Interaction, Scenario, and Display Presence Scores between low and high fidelity conditions.}
    \Description{The mean Interaction, Scenario, and Display Presence Scores between low and high fidelity conditions.}
    \label{fig:FidelityMeans}
\end{figure}

\subsection{Scenario Presence Score}
Our ANOVA for the Scenario Presence Score revealed significance for scenario fidelity level employed $F_{1, 54} = 8.6579, p < 0.001, \eta^2 = 0.138$. For high scenario fidelity conditions, we had a mean Scenario Presence Score of 5.04 with a standard deviation of 1.12. Conversely, the low scenario fidelity conditions had a mean of 4.78 and a standard deviation of 1.19 Figure \ref{fig:FidelityMeans} highlights the mean scores between the low scenario fidelity and high scenario fidelity conditions.

% Our RM-ANOVA results revealed significant main effects for interaction fidelity $F_{1, 54} = 85.46, p < 0.001, \eta^2 = 0.613$, scenario fidelity $F_{1, 54} = 9.12, p = 0.002, \eta^2 = 0.145$, and display fidelity $F_{1, 54} = 9.54, p = 0.002, \eta^2 = 0.15$. Our analysis did not reveal any significant interaction effects for the Scenario Presence Score. Following our RM-ANOVA, we conducted post-hoc analyses for each of our main effects. For interaction fidelity, we found that virtual hands led to higher Scenario Presence Scores than the ray cast interaction technique $t_{54} = 9.24, p < 0.001$. In regard to scenario fidelity, we found that consistent gravity also led to higher Scenario Presence Scores than inconsistent gravity $t_{54} = 3.02, p = 0.003$. Finally, conditions with higher display fidelity also led to higher Scenario Presence Scores than the lower display fidelity environment $t_{54} = 3.09, p = 0.002$. Figure \ref{fig:MeanScenScores} shows the mean Scenario Presence Scores from our validation study.

% \begin{figure}[]
%     \centering
%     \includegraphics[width=\linewidth]{images/MeanScenScore.png}
%     \captionsetup{justification=centering}
%     \caption{The mean Scenario Presence Scores across the 8 conditions of our validation study}
%     \label{fig:MeanScenScores}
% \end{figure}

\subsection{Display Presence Score}
Lastly, in regard to Display Presence Score, our ANOVA revealed significance for level of display fidelity $F_{1, 54} = 6.871, p < 0.001, \eta^2 = 0.113$. In high display fidelity conditions, the mean Display Presence Score was 4.58 with a standard deviation of 1.22. In low display fidelity conditions, the mean was 4.37 with a standard deviation of 1.18. Figure \ref{fig:FidelityMeans} shows the mean Display Presence Scores for the low display fidelity conditions and high display fidelity conditions.

\section{Discussion}
% All these surveys starts with an expert's opinion
% Consult a group of experts to enhance the outcomes of the presence scale

% EFA we see that loadings are different than our Delphi Study results
% Even with our questionnaire, we validated a different theoretical model

% - We don't get to claim how crucial our contribution is
% - Bridging the gap between immersion and presence (can be expanded/re-written)
%     - Describe how Sys Fidelity is an abstraction of immersion
%     - Two things:  
%         - Operationalization the slater version of immersion via Sys Fidelity than previously
%         - Provides a theoretical basis of interpreting the results of the survey
%             - You can make hypotheses based on the FPS
%             - Include examples
In this paper, we have demonstrated the capabilities that FPS provides with respect to measuring presence. Through our grounding in system fidelity, FPS ventures to strengthen the relationship between the fidelity of a VR system and presence. This is further supported by our reported results of our exploratory factor analysis and validation study. Additionally, the FPS is the result of consensus-driven design. In our Delphi study, we uncovered which questions our experts believe are the critical questions to ask when measuring presence. From our results in our Delphi study and validation study, we also present the implications of our results and future research directions.

\subsection{Bridging the Gap between System Fidelity and Presence}
% In our Delphi study, experts were asked to evaluate which component(s) of system fidelity affect the response to any given presence question. This evaluation culminated into the finalized survey, which can be found in Table \ref{tab:FPSImportanceTable}.

In previous literature, there are instances where researchers have noted the disconnect between system fidelity factors and the sense of presence, particularly in how system fidelity can impact presence. Skarbez et al. \cite{skarbez2017survey} mentions that a ``potential shortcoming of presence as a generalizable measure is that it does not account for the realism of the scenario being presented.'' Souza et al.\cite{souza2021measuring} also posits a similar thought wherein they mention a lack of clarity on whether or not there exists a relationship between the sense of presence and different technological factors.

Given this premise, we believe that FPS provides an opportunity to explicitly draw the connection between the fidelity features in a VE and the sense of presence. The system fidelity framework set by McMahan et al. \cite{McMahan2018, mcmahan2016affect} affords us the ability to meaningfully categorize immersive characteristics of a VR system. As mentioned in section 1, we describe interaction fidelity, scenario fidelity, and display fidelity individually and how they work together to form a cohesive virtual experience. Through the FPS, we can now report on those individual components of the VR system that influence presence as we have shown in section \ref{resultReport}. This can lead to better operationalization of presence as we can extend our results to discuss what components of the VE affected sense of presence and compare and contrast VEs from other studies.

The ability to discuss and break down presence measurement by system fidelity component is useful for being able to compare presence results from different environments. Jicol et al. \cite{jicol2021effects} conducted a study in which they measured the sense of presence across four environment types. In their study, they modified the environment representation, the actions a participant can engage in, and the audio being played. In another study, McMahan et al. evaluated the display fidelity and interaction fidelity in a first-person shooter environment \cite{mcmahan2012evaluating}. Inherent in their study design, McMahan et al. evaluated the impact of changing fidelity on their participants' sense of presence. In both of these examples, a tool like FPS would have been beneficial as, in addition to measuring presence, they can also report presence in respect to each of the components of system fidelity regardless of the study conditions and research questions. The capability of being able to measure presence ratings within the components of system fidelity can lead to improved future work where researchers and developers can report and compare within their own and across others' VR environments.

\subsection{Consensus-driven Survey Design}
% - A Survey made of multiple backgrounds of thought (Diversity of thought)
%     - Be careful on how we phrase it
%     - The input comes from multiple schools (5 surveys)
%     - Output is affected by 14-16 researchers
A key component to the development of FPS is that the finalized items are a result of applying the Delphi method to the presence domain. In previous work, the notion that there is not a single presence measurement tool or theoretical grounding of presence is a reoccurring theme \cite{souza2021measuring, skarbez2017survey, lee2004presence}. A possible reason for this phenomenon is that there is no single presence instrument that is developed with consensus in mind. FPS represents a significant step forward in drawing consensus across the presence research community due to our Delphi study's diversity of input and output.

In regard to diversity of input, we note early on that the items in FPS originate from WS-PQ \cite{witmer1998measuring}, IPQ \cite{schubert2001experience}, SUS \cite{usoh2000using}, MPS \cite{makransky2017development}, and SPES \cite{hartmann2015spatial}. 
These surveys represent a broad range of intellectual schools of thought and theory in the realm of presence in virtual environments for VR. They each pose their own psychological factoring and theoretical grounding of presence as seen in Table \ref{tab:RelatedWork}. One of the aims of FPS was to draw out the best qualities of these questionnaires in the form of a new presence measurement tool.

By design of the Delphi method, an additional goal of the FPS was to gain diversity of output through our experts evaluating the inventory of each of the aforementioned questionnaires. Having multiple experts afforded us the opportunity to gain decades of research experience and differing perspectives. This approach differs greatly from previous questionnaires as most questionnaires are developed under the guidance of a singular lab.

\subsection{System Fidelity and Plausibility Illusion}
With system fidelity being applied to presence, we consider the possibility that system fidelity has close ties with another construct: Plausibility Illusion. In Slater's work \cite{slater2009place}, plausibility illusion is defined as the illusion that a scenario being depicted is actually occurring. In comparison, system fidelity is the objective degree in which real-world interactions or experiences are replicated by an interactive system \cite{McMahan2018, mcmahan2016affect}. In our understanding of both of these concepts, plausibility illusion is linked more towards the subjective response a user has to the sensory stimuli present in a virtual experience \cite{slater2009place, laviola20173d} whereas system fidelity is linked to the objective design decisions a developer/researcher included in a virtual experience. In regard to the FPS, plausibility illusion influences the responses individuals report on the survey. When exposed to a virtual environment, users will determine how believable the environment is to them, which in turn affects plausibility illusion and, subsequently, sense of presence.
To further clarify, we consider the idea of a virtual environment set in a fantasy world, such as those highlighted in Rogers et al.'s literature review of realism in video games \cite{rogers2022much}  (e.g. \textit{Skyrim}). In a fantasy world, we have creatures such as centaurs, a creature with a body of a horse paired with the upper torso of a human, that can roam the environment. Inherently, centaurs do not exist in the real world and are therefore unrealistic. However, a user can still experience presence when engaged in a fantasy virtual environment. If the behaviors of the centaur match the real-world counterpart behavior and interaction (e.g. the human torso behaves like a human and the horse body behaves like a horse) it is possible for a user to experience a high level of presence and plausibility illusion. This previous case describes a high scenario fidelity experience as users would experience high physical, attribute, and behavioral coherence due to the centaur being developed based on real-world counterparts. On the converse side, if the centaur began roaming the world with flexible horse legs, similar to an octopus, we would be describing a low scenario fidelity experience as the physical coherence and attribute coherence are systematically changed.
% For example, a virtual experience developed with a four-legged monster can be perceived as an unrealistic scenario. If the monsters' behavior (e.g. walking/running properly) matches the user's expectation of how the monster would behave, then the scenario can be considered high scenario fidelity and increased level of plausibility illusion depending on their response to the FPS. However, if the monster's behavior was modified to it floating around or it not detecting walls and clipping through them, then the scenario fidelity was altered by the developer, which can lead to lowered levels of plausibility illusion.}

% \subsection{The impact of Interaction Fidelity on Presence}
% After analyzing the calculated Presence scores from our validation study, we observed that interaction fidelity is a more significant impact factor on sense of presence when compared scenario fidelity and display fidelity. Within the System fidelity framework, interaction fidelity is the first component the user acknowledges in the virtual environment. Similarly, within our study environment, participants first evaluated whether they had virtual hands or the ray casts prior to exploring the kitchen environment to put dishes away.

\subsection{Limitations and Future Work}
One of the limitations of the FPS is the potential biases within our expert participant pool. First, within the 16 expert participants, there exists the possibility of biased perspectives among those who partook in the Delphi study. As highlighted in Table \ref{tab:ParticipantChar}, our participant pool predominantly exists within North America and Europe, which may not effectively represent other cultures and perspectives. Initially, we contacted 90 potential experts with the intent of having a broad range of both intellectual and cultural perspectives. Due to factors outside of our control (e.g., scheduling), there was no feasible way to absolutely ensure diversity of culture into our Delphi study. 

Another limitation is that the FPS needs to be applied to more studies and broader contexts. We acknowledge that our validation study is in one task and context. However, in its current state, FPS is validated and ready to be used in future VR studies given our exploratory factor analysis. Future work will involve administering the FPS in a variety of environments and tasks, which will in turn contribute to a comprehensive confirmatory factor analysis in the future. In our study, we intentionally focused on extreme anchors for each fidelity component. For example, physical/attribute coherence within scenario fidelity and visual/auditory fidelity within Display Fidelity, to minimize potential confounds. Hence, we did not examine additional modifications to each of the system fidelity components, paving the path for future work to investigate the influence of these additional aspects to better generalize our findings and the application of FPS to different virtual environments.

An additional limitation is with regard to avatar embodiment in VR environments. Recent work, such as Lugrin et al.~\cite{lugrin2018} examined how VR self-body avatars influence the sense of presence through the enhancement of body ownership and spatial awareness. However, their findings suggest that these effects might not be significant all the time especially for action-based VR games, where task performance and control take precedence over self avatar representations and embodiment. In our work, we emphasize on the sense of presence within the VE itself without the inclusion of avatar embodiment. This was done to ensure that our results generalize to a variety of VR experiences (e.g. ones without self-avatar representations). We also considered the possibility that the addition of self-body avatars could introduce confounding factors and additional complexities. We acknowledge the nuanced contributions of avatar embodiment ~\cite{jung2018over, do2023valid, peck2021avatar, nowak2003effect, do2024stepping, do2024cultural} and believe there is an opportunity to apply the FPS to scenarios that leverage self-body avatars.

% One key area of future work is to expand the usage of the FPS to better understand the relationship between system fidelity and presence. Within our presented work, we have validated our study within a object manipulation context in which the components of system fidelity were systematically altered and evaluated. Future work would encompass comparing the results of the FPS to the results previously developed presence questionnaires within the same study.

\section{Conclusion}
In this paper, we have presented the development and validation of the Fidelity-based Presence Scale. In our developmental stage, we applied iterative, consensus-driven review to learn what questions experts from the VR research community deem most important. Additionally, the FPS was grounded in the system fidelity framework as we believe that future researchers and developers would greatly benefit from understanding how components of their VR applications and their fidelity impacted a user's sense of presence. The results of our Delphi study revealed a subset of 11 questions drawn from five of the most highly cited presence questionnaires for VR experiences. Following our Delphi study, we conducted a validation user study (n=55) and performed an exploratory analysis on the collected data. The results of our validation study showcased FPS' capability of identifying which component affects presence, and from our EFA, we finalized the inventory of the FPS to 10 items with their respective factor loadings. Between the Delphi study and validation methodology, we have shown the transition from theory to validation of the FPS. We also have shown how to utilize the FPS and how each component of system fidelity can be evaluated to provide additional insights and results regarding a user study. Through grounding the FPS in the system fidelity framework, our work represents a significant stride forward in presence measurement as the FPS affords the ability to inform on which components of an environment impacted the sense of presence. This capability can lead to improved comparisons between VR environments and future design implications for both developers and researchers.

\begin{acks}
The authors would like to thank Delphi Study participants (who agreed to name acknowledgment and contribution. They are listed in alphabetical order):  Dr. Sarker Asish, Dr. Sabarish Babu, Dr. Maximino Bessa, Dr. Gerd Bruder, Dr. Marcello Carrozzino, Dr. Nicola Döring, Dr. Regis Kopper, Dr. Ernst Kruijff, Dr. Miguel Melo, Dr. Anthony Steed, and Dr. Pamela Wisniewski.
\end{acks}

\bibliographystyle{ACM-Reference-Format}
\bibliography{sample-base}

\appendix

\end{document}